\documentclass[a4paper, twocolumn]{article}

\usepackage[utf8]{inputenc}
\usepackage{latexsym}
\usepackage{multirow}
\usepackage{longtable}
\usepackage{authblk}
\usepackage{comment}
\usepackage{tabularx}
\usepackage{placeins}
\usepackage{booktabs}

\usepackage{graphicx}
\usepackage{natbib}
\usepackage{amsmath}
\usepackage{amssymb}
\usepackage{subcaption} 
\usepackage{doi} 
\usepackage{xcolor}
\usepackage{parskip}
\usepackage{enumitem}
\setlist[itemize]{nosep, topsep=2pt, parsep=2pt}
\usepackage{hyperref} 
\hypersetup{
    colorlinks=true,       
    linkcolor=blue!70!black,  
    citecolor=green!50!black, 
    urlcolor=red!70!black,   
    pdftitle={Título de tu Documento},    
    pdfauthor={Tu Nombre},               
    pdfpagemode=UseNone    
}

\title{Design of nSPIRAL: a Neutron Spectrometric Position-Weighted Radial Array Long-Counter}

\author[1]{E. Mendoza}
\author[1]{J. Guerrero-C\'{a}novas}
\author[1]{D. Bachiller-Perea}
\author[1]{C. Ballesteros}
\author[1]{D. Cano-Ott}
\author[1]{J. Garc\'{i}a}
\author[1]{J. Llanes Gamonoso}
\author[1]{T. Mart\'{i}nez}
\author[1]{A.~P\'{e}rez de Rada Fiol}
\author[1]{J. Plaza}
\author[1]{L. Romero}
\author[1]{R. Santorelli}
\author[1]{D. Villamar\'{i}n}
\author[2]{O. Alonso-Sa\~{n}udo}
\author[3,4]{G. Au\~{n}\'{o}n}
\author[5]{M.J.G. Borge}
\author[2,5]{J.A. Briz}
\author[6]{F. Calvi\~{n}o}
\author[6]{G. Cort\'{e}s}
\author[6]{A. De Blas}
\author[3,4]{B. Fern\'{a}ndez}
\author[2]{L.M. Fraile}
\author[7]{G. Garc\'{i}a}
\author[6]{R. Garc\'{i}a}
\author[7]{V. Garc\'{i}a-Tavora}
\author[3,4]{C. Guerrero}
\author[5]{A. Illana}
\author[6]{N. Mont-Geli}
\author[2]{V. Mart\'{i}nez-Nouvilas}
\author[8]{E. N\'{a}cher}
\author[5]{A. Nerio-Aguirre}
\author[6]{M. Pall\`as}    
\author[5]{A. Perea}
\author[6]{D. Soler}
\author[8]{J.L. Ta\'{i}n}
\author[8]{A. Tarife\~{n}o-Saldivia}
\author[5]{O. Tengblad}
\author[2]{J.M. Ud\'{i}as
}

\affil[1]{Centro de Investigaciones Energ\'{e}ticas Medioambientales y Tecnol\'{o}gicas (CIEMAT), E-28040 Madrid, Spain}
\affil[2]{Grupo de F\'{i}sica Nuclear (GFN) and IPARCOS, Universidad Complutense de Madrid (UCM), E-28040, Madrid, Spain}
\affil[3]{Departamento de F\'{i}sica Atómica, Molecular y Nuclear, Universidad de Sevilla (US), E-41012 Sevilla, Spain}
\affil[4]{ Centro Nacional de Aceleradores (CNA) (Universidad de Sevilla (US) - Junta de Andaluc\'{i}a - CSIC), E-41092, Sevilla, Spain}
\affil[5]{Instituto de Estructura de la Materia (IEM), CSIC, E-28006 Madrid, Spain}
\affil[6]{Institut de T\`ecniques Energ\`etiques (INTE), Universitat Polit\`ecnica de Catalunya (UPC), E-08028, Barcelona, Spain}
\affil[7]{Centro de Micro-An\'{a}lisis de Materiales (CMAM), Universidad Aut\'{o}noma de Madrid (UAM), E-28049, Madrid, Spain}
\affil[8]{  Instituto de F\'{i}sica Corpuscular (IFIC), CSIC – Univ. Valencia (UV), E-46071, Valencia, Spain}

\begin{document}

\maketitle

\begin{abstract}

We present here the design of a neutron long counter intended to measure neutrons originating from radioactive decays or nuclear reactions. Like other similar detectors, it relies on a moderator (polyethylene) and $^{3}$He tubes. A novel feature of our design is the use of the weighting technique to achieve an efficiency that is virtually independent of the energy spectrum of the neutrons being measured; a property known as \textit{flat efficiency}. This approach offers two primary advantages: it yields a flatter efficiency response than conventional analysis methods, and it provides greater flexibility regarding the geometry of the long counter. The latter has enabled us to design the detector with almost every $^{3}$He tube positioned at a different distance from the neutron source, which allows us to perform not only neutron counting but also neutron spectroscopy.

\end{abstract}

\section{Introduction} \label{sec:Introduction}

Neutron long counters have been used widely for decades to detect neutrons emitted in radioactive decays~\cite{Caballero-Folch_2018,Hall_2021,Ong_2024} or in nuclear reactions such as (p,n), ($\gamma$,n) or ($\alpha$,n)~\cite{Hansen_1967,Zyskind_1979,Brandenburg_2023,Zhou_2025}. They are made of a moderator matrix, typically polyethylene but also graphite, in which neutron detectors such as $^{3}$He or BF$_{3}$ tubes are inserted. Neutrons are generated inside this matrix and are detected by the $^{3}$He or BF$_{3}$ tubes after being moderated. They are designed to have a detection efficiency that does not depend significantly on the initial energy of the neutrons being measured (this is referred to as \textit{flat efficiency}). In this way, they are capable of counting the neutrons generated within them, regardless of their energy spectrum, which is often unknown. Examples of long neutron counters are BRIKEN~\cite{Tarifeno-Saldivia_2017,Tolosa-Delgado_2019}, NERO~\cite{Pereira_2010}, BELEN~\cite{Agramunt_2016}, miniBELEN~\cite{MontGeli_2026}, LOENIE~\cite{Mathieu_2012}, LUNA~\cite{Csedreki_2021}, TETRA~\cite{Testov_2016} and HeBGB~\cite{Brandenburg_2022}, all of them using polyethylene; and the Macklin Graphite Sphere Detector~\cite{Macklin_1957}, using graphite.

\begin{figure}[htb]
\begin{center}
\includegraphics[width=0.99\linewidth]{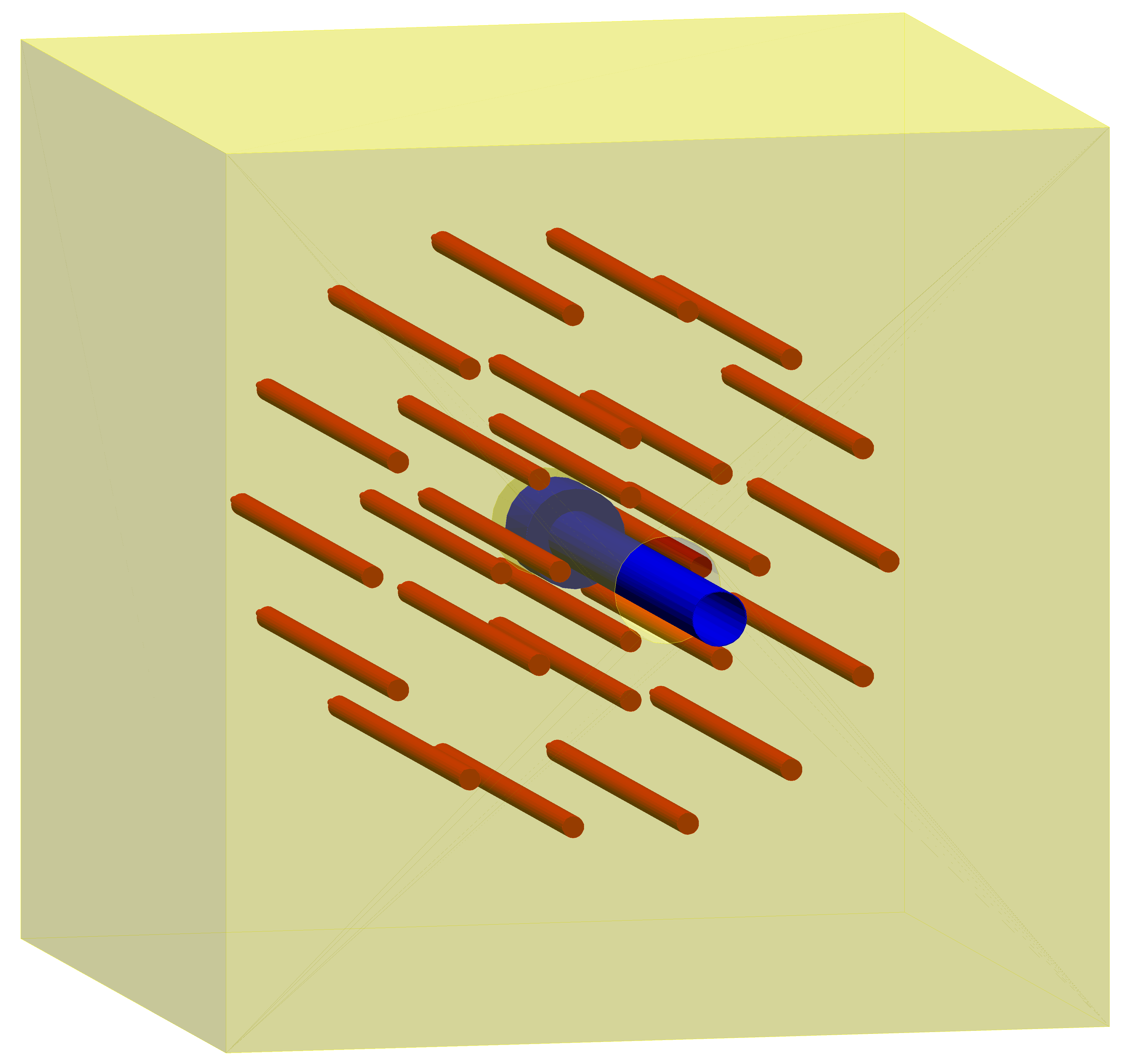} \\
\end{center}
\caption{Schematic view of a neutron long counter. It consists of a 1 m polyethylene cube containing 26 $^{3}$He tubes, arranged in 3 concentric rings. The polyethylene matrix has a central hole through which a beam tube is inserted (in blue), which carries the radioactive nuclei or incident particles to the center. Fig.~\ref{fig:Conf-3R_02} provides a front view of it.}
\label{fig:Conf-3R_01}
\end{figure}

Here, we will refer exclusively to long neutron counters that detect neutrons generated within their own interior, such as the one shown in Fig.~\ref{fig:Conf-3R_01}. There are also other detectors with the same name and similar characteristics designed to count neutrons coming from the outside, that are primarily used to measure neutron fluxes~\cite{East_1969,Lacoste_2010,Wang_2015}.

In the neutron long counters referred to here the flat-efficiency response is achieved by specifically engineering the device's geometry to produce this characteristic. This is accomplished through the careful design of the moderator dimensions, the placement of the $^{3}$He or BF$_{3}$ tubes within it, and the detection efficiency of the tubes themselves. The typical geometric configuration involves placing several parallel detector rings, each at a certain distance from the center. Those closer to the center are more efficient at detecting lower-energy neutrons, while those farther away detect higher-energy neutrons. By adjusting the distances of the rings from the center, the number of tubes in each ring, and the efficiencies of each tube, the detection system becomes less dependent on the neutron energy. To vary the detection efficiency of the tubes, different sizes, different gas pressures, or even partial coatings of cadmium, which absorbs thermal neutrons, are used.

In this paper, we present the design of a new long counter type neutron detector. Instead of adjusting the detector geometry, we propose using the weighting technique as a tool to achieve a flat-efficiency response of our detection device. This technique consists of multiplying the number of counts recorded in each tube by a value, referred to as a \textit{weight}, to estimate the number of neutrons emitted from the inside of the long counter. We have not found that the weighting technique is used in the context of long counters, but it is used in other areas~\cite{Macklin_1967,Abbondanno_2004,Mendoza_2023}.

As we shall see, the weighting technique allows for flatter efficiency responses. Furthermore, the geometric arrangement of the tubes within the moderator becomes much more flexible. This has enabled us to position the tubes at varying distances from the source, allowing us not only to count the number of neutrons but also to obtain information about their energy spectrum.

This article is structured as follows. Sections~\ref{sec:Weighting_Technique},~\ref{sec:Efficiency_Flatness}, and~\ref{sec:Weight_Fitting} are dedicated to describing the weighting technique: Section~\ref{sec:Weighting_Technique} describes the technique; Section~\ref{sec:Efficiency_Flatness} defines the parameters that quantify the flatness of the efficiency; and Section~\ref{sec:Weight_Fitting} addresses how to obtain the appropriate weights, given a specific geometric configuration of the detection device. In Section~\ref{sec:AdjGeom}, we present the optimization process for the geometry of a long counter consisting of three concentric rings of $^{3}$He tubes, a configuration similar to many such long counters currently in use. Sections~\ref{sec:Design_nSPIRAL} and~\ref{sec:Performance_nSPIRAL} are dedicated to the design and performance of our proposed long counter, which we have named nSPIRAL (neutron Spectrometric Position-weIghted Radial Array Long-counter). We first present an intermediate design, in which the $^{3}$He tubes are arranged in a spiral (i.e., covering 4$\pi$); we then move on to the final design, in which the entire spiral is positioned on one side of the neutron source (i.e., covering 2$\pi$), with the goal of reducing the size of the long counter. The optimized three-ring-geometry detector described in Section~\ref{sec:AdjGeom} allows for a comparison between the performance of our final design and that of a detector with a more conventional geometry. Finally, Section~\ref{sec:Conclusions} provides a summary and conclusions.

\section{The Weighting Technique} \label{sec:Weighting_Technique}

Let us assume that we have a detection device made of $N$ individual detectors. This detector device is designed to measure the number of neutrons emitted from a certain source $S$ in a period of time. Let us also assume that each source neutron can be detected by at most one of the individual detectors. In other words, for each source neutron, we can have one count in one of the detectors, or none in any of them; i.e., a neutron cannot result in two counts. Let us also assume that the only information the detectors provide is that a neutron has been detected; that is, there is no information about its energy, position, or any other parameter.

Let us now define $\varepsilon_{S,i}$ as the probability that a neutron from the source $S$ will be detected by one of the detectors, the one labeled with the letter $i$ ($i$ therefore takes values between 1 and $N$). This variable $\varepsilon_{S,i}$ is commonly known as the detection efficiency of detector $i$ (for the source $S$). Let us imagine that we perform a measurement over a period of time, in which each detector records a number of counts $C_{i}$. A common approach would then be to estimate the number of neutrons emitted by the source ($N_{s}$) as:

\begin{equation} \label{eq:Estimator1}
\hat{N}_{s}=\frac{\sum_{i}C_{i}}{\sum_{i}\varepsilon_{S,i}}=\frac{\sum_{i}C_{i}}{\varepsilon_{S}}
\end{equation}
where the hat indicates that we are referring to an estimator, i.e., $N_{s}$ is the number of neutrons emitted by the source and $\hat{N}_{s}$ is an (unbiased and consistent~\cite{Wackerly_2008}) estimator of $N_{s}$. Here we would normally define  $\varepsilon_{S}=\sum_{i}\varepsilon_{S,i}$ as the detection efficiency of the detection device for the neutron source $S$.

In general, detection efficiencies $\varepsilon_{S}$ and $\varepsilon_{S,i}$ will depend on the characteristics of the source $S$, and specifically on the distribution in energy, direction and position of the source neutrons. If the characteristics of the sources are known, then $\varepsilon_{S}$, which is needed to compute $\hat{N}_{s}$, could in principle be obtained from Monte Carlo calculations. However, what often happens is that the characteristics of the source are not known precisely.

In the case of neutrons emitted in radioactive decays, for example, the position is often known and the source is assumed to be isotropic (although if measured in coincidence with a $\beta$ detector this is not so clear), but the energy distribution of the neutrons is unknown. To deal with this situation the detection device is then designed so that $\varepsilon_{S}=\sum_{i}\varepsilon_{S,i}$ is only minimally dependent on the energy distribution of the source neutrons. To achieve this, the fundamental task is to find a geometric configuration of the $^{3}$He or BF$_{3}$ tubes within the moderator matrix that satisfies the aforementioned condition~\cite{Tarifeno-Saldivia_2017,Mathieu_2012,Brandenburg_2022,MontGeli_2023}.

What we propose here is, in addition to seeking an optimal geometry, to use the weighting technique to achieve a flat-efficiency response. With a flat-efficiency response we mean that the efficiency of the detection device does not depend much on those characteristics of the source that we do not know, such as the energy distribution and, potentially, angular and spatial distributions as well.

The weighting technique involves assigning a weight to the counts recorded by the individual detectors (the $^{3}$He or BF$_{3}$ tubes). These weights vary from one detector to another. The estimator in Eq.~\ref{eq:Estimator1} is a possible estimator of $N_{s}$, but it is not the only one. Any expression of the form:

\begin{equation} \label{eq:Estimator2}
\hat{N}_{s}=\frac{\sum_{i}w_{i}\cdot C_{i}}{\sum_{i}w_{i}\cdot\varepsilon_{S,i}}
\end{equation}
, where $w_{i}$ (the weights) are any real numbers such that $\sum_{i}w_{i}\cdot\varepsilon_{S,i}\neq0$, is also an (unbiased and consistent) estimator of $N_{s}$. In an intuitive way, it is easy to see that as $C_{i}$ increase, they approach $N_{s}\cdot \varepsilon_{S,i}$; and consequently $\hat{N}_{s}$ approaches $N_{s}$.

By analogy with Eq.~\ref{eq:Estimator1}, we can define $\varepsilon^{w}_{S}=\sum_{i}w_{i}\cdot\varepsilon_{S,i}$. Thus, when using the weighting technique, the flat-efficiency response of the detection device will be achieved if $\varepsilon^{w}_{S}$ does not depend much on the characteristics of the source that are not known. The advantage of the weighting technique is that to achieve a flat-efficiency response we can adjust the values of $w_{i}$ in addition to the geometric configuration of the detection device.

Regarding the uncertainty due to counting statistics, if we assume that $\Delta C_{i}=\sqrt{C_{i}}$, then we have that:

\begin{equation} \label{eq:UncEstimator2_1}
\Delta \hat{N}_{s}=\frac{\sqrt{\sum_{i}w_{i}^{2}\cdot C_{i}}}{\sum_{i}w_{i}\cdot\varepsilon_{S,i}}
\end{equation}
and hence:
\begin{equation} \label{eq:UncEstimator2_2}
\frac{\Delta \hat{N}_{s}}{\hat{N}_{s}}=\frac{\sqrt{\sum_{i}w_{i}^{2}\cdot C_{i}}}{\sum_{i}w_{i}\cdot C_{i}}
\end{equation}

It can be demonstrated that the expression in Eq.~\ref{eq:UncEstimator2_2} has a minimum when all $w_{i}$ have the same value. In this particular case Eq.~\ref{eq:Estimator2} becomes Eq.~\ref{eq:Estimator1}, which is the one normally used because it converges faster. In other words, for a given number of counts $C_{i}$, if we use Eq.~\ref{eq:UncEstimator2_2} with all $w_{i}=1$, i.e. Eq.~\ref{eq:Estimator1}, we will obtain an uncertainty due to counting statistics in $\hat{N}_{s}$ smaller than if we use weights $w_{i}$ with different values from each other. Furthermore, the more different the values of $w_{i}$ are from each other, the greater the uncertainty due to counting statistics. However, the aim of using the weighting technique is to partially sacrifice the convergence speed of $\hat{N}_{s}$ in order to achieve a flatter-efficiency.

Continuing our analysis of the convergence speed of $\hat{N}_{s}$, we can see that as the values of $C_{i}$ increase and converge to $N_{s}\cdot \varepsilon_{S,i}$,  Eq.~\ref{eq:UncEstimator2_2} approaches:

\begin{equation} \label{eq:UncEstimator2_3}
\frac{\Delta \hat{N}_{s}}{\hat{N}_{s}} \simeq
\frac{1}{\sqrt{N_{s}}} \frac{\sqrt{\sum_{i}w_{i}^{2}\cdot \varepsilon_{S,i}}}{\sum_{i}w_{i}\cdot \varepsilon_{S,i}}=\frac{1}{\sqrt{N_{s} \varepsilon^{*}_{S}}}
\end{equation}
where
\begin{equation} \label{eq:EquivEff}
 \varepsilon^{*}_{S}=\frac{(\sum_{i}w_{i}\cdot \varepsilon_{S,i})^2}{\sum_{i}w_{i}^{2}\cdot \varepsilon_{S,i}}
\end{equation}

When the weighting technique is not used, it is well known that as the number of registered counts increase, $\frac{\Delta \hat{N}_{s}}{\hat{N}_{s}}$ approaches:

\begin{equation} \label{eq:UncEstimator1_2}
\frac{\Delta \hat{N}_{s}}{\hat{N}_{s}}=\frac{1}{\sqrt{\sum{C_{i}}}}\simeq\frac{1}{\sqrt{N_{s} \varepsilon_{S}}}
\end{equation}

Comparing Eq.~\ref{eq:UncEstimator2_3} and Eq.~\ref{eq:UncEstimator1_2}, we see that when using the weighting technique, the quantity $\varepsilon^{*}_{S}$ plays the same role as the efficiency $\varepsilon_{S}$ when not using the weighting technique in determining the uncertainties due to counting statistics.

Therefore, when we use the weighting technique, instead of one, we have three types of detection efficiencies:
\begin{itemize}
 \item $\varepsilon_{S}=\sum{\varepsilon_{S,i}}$, which is the (standard) efficiency of the detection device. It can be used to estimate the expected number of counts we will have after a certain number of neutrons $N_{s}$ have been emitted by the source. 
 \item $\varepsilon^{w}_{S}=\sum_{i}w_{i}\cdot\varepsilon_{S,i}$, which is the \textit{efficiency} which has to be considered to achieve a flat-efficiency (Eq.~\ref{eq:Estimator2}).
 \item $\varepsilon^{*}_{S}$, defined in Eq.~\ref{eq:EquivEff}, which is the \textit{efficiency} which plays the same role as the efficiency when not using the weighting technique in determining the uncertainties due to counting statistics. For a certain number of source neutrons $N_{S}$, the larger the value of $\varepsilon^{*}_{S}$, the smaller the uncertainties due to counting statistics.
\end{itemize}

When not using the weighting technique these three \textit{efficiencies} are the same.

\section{Quantifying Efficiency Flatness} \label{sec:Efficiency_Flatness}

Efficiency flatness is achieved when $\varepsilon^{w}_{S}$ is largely independent of the source variables that are unknown a priori. To accomplish this, we define a parameter to quantify efficiency flatness regarding these unknown variables and subsequently optimize the geometry.

The literature has focused exclusively on neutron energy as an unknown variable, as is the case with the work presented here. However, a more general approach should consider energy ($E_{n}$), the polar ($\theta$) and azimuthal ($\phi$) angles, as well as position ($\vec{x}$). These can be grouped into a single variable $\vec{\lambda}$, representing the source parameters with unknown probability distributions. For instance, an isotropic source with a known position but an unknown energy distribution yields $\vec{\lambda}=\{E_{n}\}$. Conversely, if both the energy and angular distributions relative to a specific direction (such as the incident alpha beam in ($\alpha$,n) measurements) are unknown, then $\vec{\lambda}=\{E_{n},\theta\}$. In the most general case, $\vec{\lambda}=\{E_{n},\theta,\phi,\vec{x}\}$.

Let $\varepsilon_{i}(\vec{\lambda})$ denote the probability of detecting a neutron emitted with a specific energy, angles ... given by $\vec{\lambda}$, for the detector labeled with the letter $i$. Each neutron source to be considered will be defined by a probability distribution $p_{S}(\vec{\lambda})$. Now we can discretize the $\vec{\lambda}$-space into a certain number of intervals (or n-intervals where n is the dimension of the $\vec{\lambda}$-space), and we can define $\varepsilon_{i,j}$ and $p_{S,j}$ respectively as the average value of $\varepsilon_{i}(\vec{\lambda})$ and $p_{S}(\vec{\lambda})$ in the interval labeled with the letter $j$. Thus, if, for example, $\vec{\lambda}=\{E_{n}\}$, then the values of $p_{S,j}$ will define the energy distribution of the source $S$, which has been discretized in a certain number of intervals (i.e. as a one-dimensional histogram); if $\vec{\lambda}=\{E_{n},\theta\}$, then the 2-intervals (rectangles) will be in $E_{n}$ and in $\theta$ (i.e. as a two-dimensional histogram); and so on. In both cases, $\varepsilon_{i,j}$ will be the average efficiency of the tube $i$ of detecting neutrons emitted with the characteristics provided by the interval or rectangle labeled with the letter $j$. $\varepsilon_{i,j}$ is also commonly referred to as the response matrix of the detection device.

We will then have that:

\begin{equation} \label{eq:w-Eff-01}
\varepsilon_{S,i} \simeq \sum_{j}\varepsilon_{i,j}\cdot p_{S,j}
\end{equation}
and
\begin{equation} \label{eq:w-Eff-02}
\varepsilon^{w}_{S}=\sum_{i}w_{i}\cdot\varepsilon_{S,i}\simeq \sum_{i,j}w_{i}\cdot\varepsilon_{i,j}\cdot p_{S,j}=\sum_{j}y_{j}\cdot p_{S,j} \\
\end{equation}
with:
\begin{equation} \label{eq:w-Eff-03}
y_{j}=\sum_{i}w_{i}\cdot\varepsilon_{i,j}
\end{equation}

Recall that our goal is for $\varepsilon^{w}_{S}$ to depend little on $p_{S}(\vec{\lambda})$. From Eq.~\ref{eq:w-Eff-02} this will be achieved if the values of $y_{j}=\sum_{i}w_{i}\cdot\varepsilon_{i,j}$ take similar values for all the $j$-intervals. The more similar the values of $y_{j}$ are, the flatter the efficiency will be.

To quantify this, we define a parameter which indicates how flat the efficiency of the detection device is. In previous works~\cite{MontGeli_2026} the \textit{flatness factor} $F$, was defined as the ratio between the maximum and minimum values of $y_{j}$:

\begin{equation} \label{eq:Flatness1}
F_{1}= \frac{\max_j \{y_j\}}{\min_j \{y_j\}}
\end{equation}

However, in this work we prefer to use:

\begin{equation} \label{eq:Flatness2}
F_{2}= \frac{1}{M}\sum_{j=1}^{M}\left(\frac{y_{j}-y_{m}}{y_{m}} \right)^{2}
\end{equation}
where $M$ is the number of $j$-intervals and $y_{m}$ is the mean value of the $y_{j}$. The smaller the value of $F_{2}$, the flatter the efficiency.

It is also worth adding that using this discretization of the source, the value of $\varepsilon^{*}_{S}$ is computed as (Eqs.~\ref{eq:EquivEff} and~\ref{eq:w-Eff-01}):

\begin{equation} \label{eq:EquivEff-02}
 \varepsilon^{*}_{S}\simeq\frac{(\sum_{i,j}w_{i}\cdot \varepsilon_{i,j}\cdot p_{S,j})^2}{\sum_{i,j}w_{i}^{2}\cdot \varepsilon_{i,j}\cdot p_{S,j}}
\end{equation}

\section{Weight Fitting Procedure} \label{sec:Weight_Fitting}

Based on what we have discussed in the previous sections, we want our detection system to have the lowest possible $F_{2}$ value, so the detection device has a flatter efficiency with respect to the source variables that are unknown a priori, and the highest possible $\varepsilon^{*}_{S}$ value, as this will minimize uncertainties due to counting statistics. We can then define a figure of merit $FOM$, as the parameter to be minimized, as follows:

\begin{equation} \label{eq:FOM}
FOM=\frac{F_{2}^{\alpha}}{\varepsilon^{*}_{S}}
\end{equation}
where $\alpha$ is a positive real number that serves to modulate the importance that minimizing $F_{2}$ or maximizing $\varepsilon^{*}_{S}$ has for us. The larger the value of $\alpha$, the more weight $F_{2}$ will have when minimizing the $FOM$, and the smaller it is, the more weight $\varepsilon^{*}_{S}$ will have.

For a certain detector device, we will then have to find the weights $w_{i}$ that minimize the $FOM$. For this, we can see from Eqs.~\ref{eq:w-Eff-03},~\ref{eq:Flatness2} and~\ref{eq:EquivEff-02} that to compute the $FOM$ for a certain set of weights $w_{i}$ we need to know the $\varepsilon_{i,j}$ values and to establish or define some $p_{S,j}$ values. This is because $F_{2}$ does not depend on the $p_{S,j}$ values, but $\varepsilon^{*}_{S}$ does. In consequence, some $p_{S}(\vec{\lambda})$ has to be assumed for calculating the $FOM$. In this work, we always consider $p_{S}(\vec{\lambda})=p_{S}(E_{n})$, and to calculate the FOM, we have assumed uniform distributions between 0 and a maximum energy, $E_{n,max}$.

In order to calculate the values of the response matrix $\varepsilon_{i,j}$, we have performed Monte Carlo simulations using the Geant4 code~\cite{Geant4_2003,Geant4_2016}. Specifically, version 11.3.2 (released in April 2025), which already contains the recent improvements in the thermal neutron treatment reported in~\cite{Thulliez_2022}.

Once the values of $\varepsilon_{i,j}$ are obtained, which will depend on the specific detection system, i.e., on its geometry, we can calculate the value of the $FOM$ for each set of weights. In other words, once we fix $\varepsilon_{i,j}$ and $p_{S,j}$, we have that $FOM=FOM(w_{1},w_{2},...,w_{N})$. For fitting the weights to minimize the $FOM$, we used the version of Minuit2~\cite{Minuit2_2005} implemented in ROOT~\cite{root_1997}, which is widely used and which we have verified works very well.

\begin{figure}[htb]
\begin{center}
\includegraphics[width=0.99\linewidth]{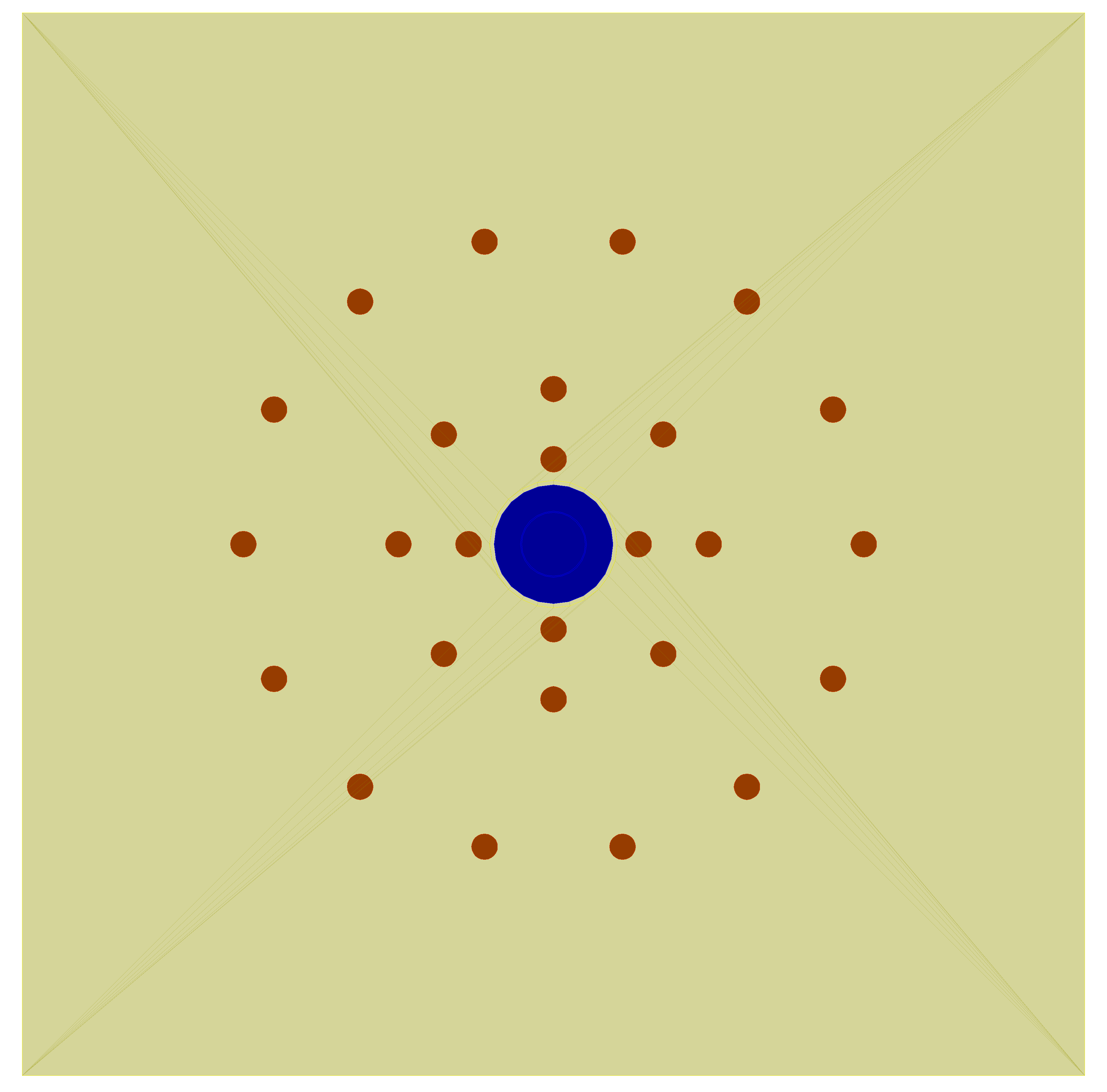} \\
\end{center}
\caption{Schematic view of a neutron long counter with 26 $^{3}$He tubes arranged in three concentric rings. The radii of the three rings have been optimized as described in Section~\ref{sec:AdjGeom}. The same geometry, shown from a different perspective, is presented in Fig.~\ref{fig:Conf-3R_01}.}
\label{fig:Conf-3R_02}
\end{figure}

To illustrate all this, we will apply it to a specific case: designing a long counter similar to those currently in use, but employing the weighting technique. We will take as moderator a polyethylene cube with sides of 1 m, containing a central hole 12 cm in diameter and 60 cm deep, necessary for the incident particle beam or radioactive nuclei. Inserted in the polyethylene matrix, we place 26 $^{3}$He tubes arranged in three concentric rings. The innermost ring consists of 4 detectors and is located at a distance d$_{1}=8.00$ cm from the center; the middle ring consists of 8 detectors and is located at a distance d$_{2}=14.6$ cm from the center; and the outermost ring consists of 14 detectors and is located at a distance d$_{3}=29.2$ cm from the center. The $^{3}$He tubes correspond to those we have in our CIEMAT group and with which we have previously worked~\cite{Plaza_2023}; specifically, those manufactured by LND Inc.~\cite{LNDINC_website}, model 252266. They are approximately 64 cm long and have a radius of 1.22 cm, with a pressure of 10 bar. The active volume is slightly smaller: 60 cm long and with a radius of 1.17 cm. We show a schematic of this arrangement in Figs.~\ref{fig:Conf-3R_01} and~\ref{fig:Conf-3R_02}. The values of the distances d$_{1}$, d$_{2}$ and d$_{3}$ have been optimized as described in Section~\ref{sec:AdjGeom}.

We have assumed that the source is a point source, located at the center of the polyethylene cube, and that is isotropic. What is unknown is the energy distribution of the source neutrons. Thus, we need the value of $\varepsilon^{w}_{S}$ of our detection device to depend as little as possible on $p_{S}(E_{n})$. To do this, we discretize the energy of the source neutrons into a certain number of intervals, and then perform simulations with Geant4 to obtain the values of $\varepsilon_{i,j}$, where the index $i$ refers to the $^{3}$He tube ($i=1,...,26$) and the index $j$ refers to the energy interval of the source neutrons. Each $\varepsilon_{i,j}$ value has been obtained by performing Geant4 simulations with an energy uniformly distributed in the interval provided by the index $j$. In this work, we have adopted linear energy $j$-intervals with a width of 100 keV. The reason for using linear intervals is that, in radioactive decays and nuclear reactions, the energies of the emitted neutrons exhibit probability distributions that are best expressed on linear rather than logarithmic scales. This is because these energies are closely related to mass differences between nuclei. Logarithmic scales are more appropriate when neutrons are moderated after undergoing elastic collisions.

Due to the symmetry of our detection device, it is expected that the $\varepsilon_{i,j}$ values will be very similar for the detectors of the same ring, so it is reasonable to associate the same weight to them. Thus, we will have only three weights $w_{1}$, $w_{2}$ and $w_{3}$ corresponding, respectively to the inner, middle and outer rings. Similarly, we can refer to the detection efficiencies of each of those rings.

\begin{figure}[htb]
\begin{center}
\includegraphics[width=0.99\linewidth]{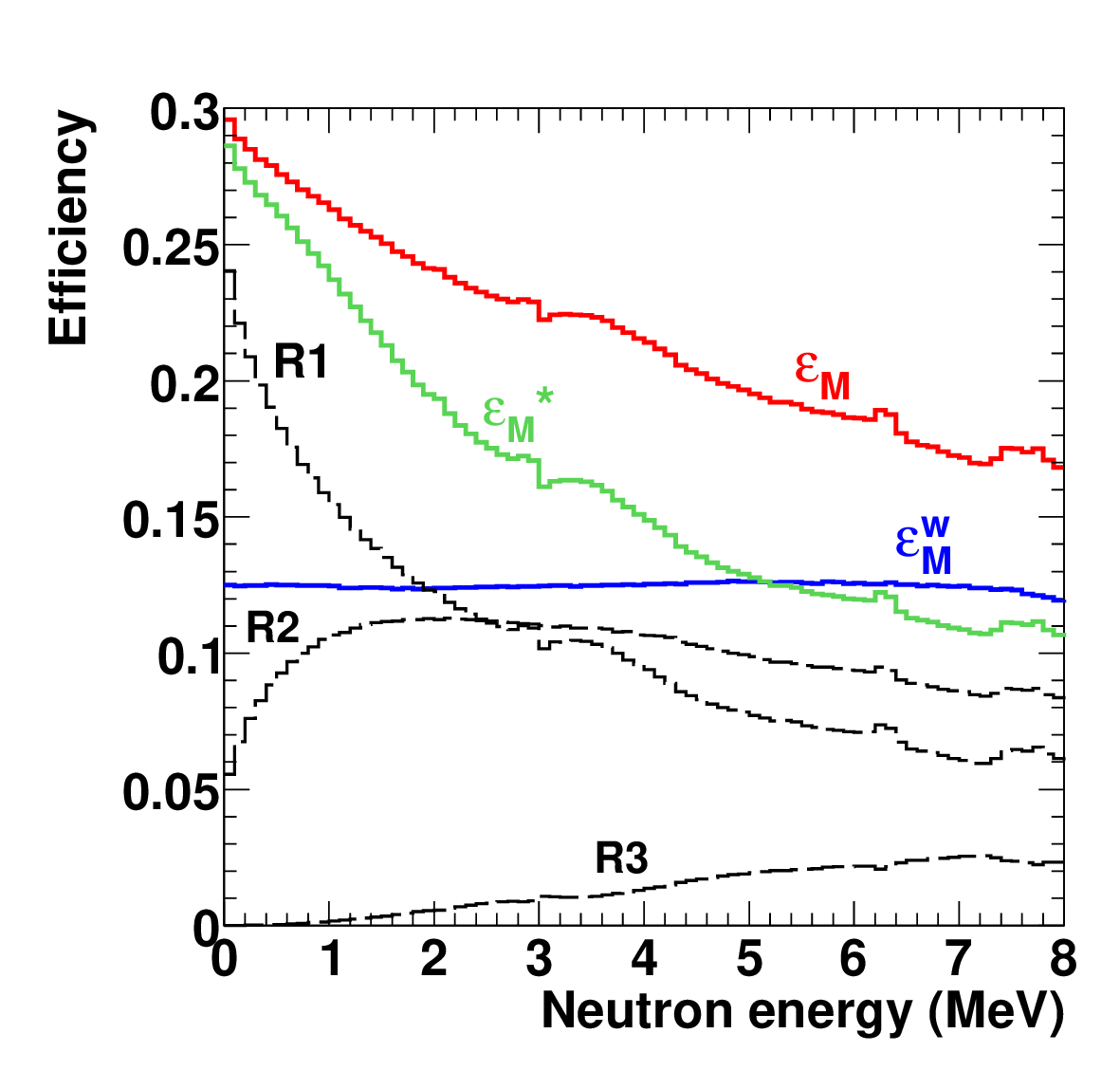} \\
\end{center}
\caption{Different efficiencies of the detection device shown in Figs.~\ref{fig:Conf-3R_01} and~\ref{fig:Conf-3R_02}, for weight values of $w_{1}=0.387$, $w_{2}=0.576$ and $w_{3}=2.04$. Details are given in the body text.}
\label{fig:EffExample01}
\end{figure}

Calculating the FOM requires defining a specific energy range of interest, which we have initially set from 0 to 8 MeV. Our initial measurements will focus on reactions induced by $\alpha$-particles from the Th and U decay chains, where $\alpha$ energies reach approximately 9 MeV. For most target nuclei, neutrons produced by 9 MeV $\alpha$-particles do not exceed 8 MeV, with some exceptions such as $^{9}$Be. If we take then $E_{n,max}=8$ MeV and we minimize the $FOM=FOM(w_{1},w_{2},w_{3})$ (with Minuit2) using an $\alpha$-value of $\alpha=0.5$, then we obtain $w_{1}=0.387$, $w_{2}=0.576$ and $w_{3}=2.04$. The obtained efficiencies using these weights are presented in Fig.~\ref{fig:EffExample01}. There we show the detection efficiencies of each of the rings as a function of the source neutron energy: these are the black curves labeled R1 (inner ring), R2 (middle ring), and R3 (outer ring). The sum of the three gives us the red curve, which is the efficiency $\varepsilon_{S}$  (conventionally defined) of the detection system. As can be seen, this efficiency varies considerably with the neutron energy. By weighting the efficiencies of the three rings with the weights reported before, we obtain the blue curve. This is the efficiency $\varepsilon^{w}_{S}=\sum_{i}w_{i}\cdot\varepsilon_{S,i}$ we want to be flat (Eq.~\ref{eq:Estimator2}). Finally, we also show the third efficiency of the detection device defined in this work, $\varepsilon^{*}_{S}$, in green, which is obtained from Eq.~\ref{eq:EquivEff-02} when using a monochromatic neutron source with energy given by the x-axis. This is the reason why we labeled $\varepsilon_{S}$, $\varepsilon^{w}_{S}$ and $\varepsilon^{*}_{S}$ as $\varepsilon_{M}$, $\varepsilon^{w}_{M}$ and $\varepsilon^{*}_{M}$, to indicate that the source $S$ is monochromatic ($M$).

\begin{table*}[ht]
\centering\
\begin{tabular}{ccccc}
\toprule
$\alpha$ & $\{w_{1},w_{2},w_{3}\}$ & $F_{1}$ & $F_{2}$ & $\varepsilon^{*}_{S}$ \\
\midrule
0.001  &  \{0.997,0.999,1.003\}  & 1.77 & $2.61\cdot10^{-2}$ &21.6\% \\
0.01   &   \{0.971,0.994,1.035\} & 1.74 & $2.48\cdot10^{-2}$ &21.6\% \\
0.08   &   \{0.706,0.930,1.364\} & 1.45 & $1.17\cdot10^{-2}$ &20.7\% \\
0.09   &   \{0.387,0.602,2.011\} & 1.06 & $1.30\cdot10^{-4}$ &15.1\% \\
0.5    &   \{0.387,0.576,2.037\} & 1.06 & $1.14\cdot10^{-4}$ &14.9\% \\
1.0    &   \{0.387,0.574,2.040\} & 1.06 & $1.14\cdot10^{-4}$ &14.8\% \\
\bottomrule
\end{tabular}
\caption{Results of fitting the weights when minimizing the $FOM$ for different values for the $\alpha$ parameter (Eq.~\ref{eq:FOM}), together with the obtained flatness factors $F_{1}$ and $F_{2}$, and $\varepsilon^{*}_{S}$. In all cases $E_{n,max}=8$ MeV.}
\label{tab:table01}
\end{table*}

\begin{table*}[ht]
\centering\
\begin{tabular}{ccccc}
\toprule
$E_{n,max}$ & $\{w_{1},w_{2},w_{3}\}$ & $F_{1}$ & $F_{2}$ & $\varepsilon^{*}_{S}$ \\
\midrule
2 MeV  &  \{0.358,0.536,2.106\} & 1.007 & $3.49\cdot10^{-6}$ &22.8\% \\
4 MeV  &  \{0.392,0.615,1.993\} & 1.010 & $7.55\cdot10^{-6}$ &19.8\% \\
6 MeV  &  \{0.394,0.620,1.986\} & 1.011 & $8.12\cdot10^{-6}$ &17.2\% \\
8 MeV  &  \{0.387,0.576,2.037\} & 1.059 & $1.14\cdot10^{-4}$ &14.9\% \\
10 MeV &  \{0.377,0.502,2.121\} & 1.126 & $8.82\cdot10^{-4}$ &12.4\% \\
\bottomrule
\end{tabular}
\caption{Results of fitting the weights when minimizing the $FOM$ for different values of the maximum energy considered for the fit, $E_{n,max}$, together with the obtained flatness factors $F_{1}$ and $F_{2}$, and $\varepsilon^{*}_{S}$. In all cases $\alpha=0.5$.}
\label{tab:table02}
\end{table*}

As can be seen in Fig.~\ref{fig:EffExample01}, we have succeeded in making our detection system rather independent of the energy distribution of the source neutrons. The values of $\varepsilon^{w}_{M}$ deviate on average by less than 0.8\% from the mean; the deviations are always less than 1.6\% from the mean for energies below 7 MeV and deviate slightly more between 7 and 8 MeV, without reaching 5\%.

Regarding the value of $\varepsilon^{*}_{S}$, if we had a uniformly distributed source the value would be $\varepsilon^{*}_{S}=14.9\%$. However, given the decreasing trend of $\varepsilon^{*}_{M}$ with the neutron energy, the lower the energy of the neutrons from the source, the higher the value of $\varepsilon^{*}_{S}$ will be. It is worth noting here that, due to the definition of $\varepsilon^{*}_{S}$ (Eq.~\ref{eq:EquivEff}), the value of $\varepsilon^{*}_{S}$ cannot be calculated by multiplying curve $\varepsilon^{*}_{M}$ by the neutron spectrum of the source, unlike $\varepsilon_{S}$.

Finally, we show in Tables~\ref{tab:table01} and~\ref{tab:table02} the results of the fits when we vary the parameter $\alpha$ (Eq.~\ref{eq:FOM}) and the maximum energy considered for the fit, $E_{n,max}$. As expected, for small values of $\alpha$, the fit prioritizes maximizing $\varepsilon^{*}_{S}$, whereas for larger values, it prioritizes minimizing $F_{2}$. In this specific case, there is a rather abrupt transition between  $\alpha=0.08$ and $\alpha=0.09$; however, in other cases we have studied, this abrupt transition occurs between different values of $\alpha$. As for $E_{n,max}$, the larger its value, the more difficult it becomes to achieve a flat efficiency.

\section{Geometry adjustment} \label{sec:AdjGeom}

In the previous section we have shown how to fit the weights $w_{i}$ to get the smallest value for the $FOM$, once defined the geometry of the detection device (i.e. $\varepsilon_{i,j}$) and the energy range of the neutrons that are to be detected (i.e. $E_{n,max}$). This would give us a $FOM$ value associated with each geometry, which would, in principle, allow comparing two geometries, provided that the only criterion for deciding which of the two is preferable is based on the one with the smallest $FOM$ value. We could then parameterize the geometry of the detection device according to some variables. In this way, the $FOM$ value would depend on these variables, and finding the best geometry would become again a problem of minimizing the $FOM$.

This is what we have done with the three-ring long counter presented in the previous section. We have parameterized the geometry using the three distances from the rings to the center of the geometry: $d_{1}$ (inner ring), $d_{2}$ (middle ring) and $d_{3}$ (outer ring); so that $FOM=FOM(d_{1},d_{2},d_{3})$.

\begin{figure*}[htb!]
\begin{center}$
\begin{array}{c c}
\includegraphics[width=0.45\linewidth]{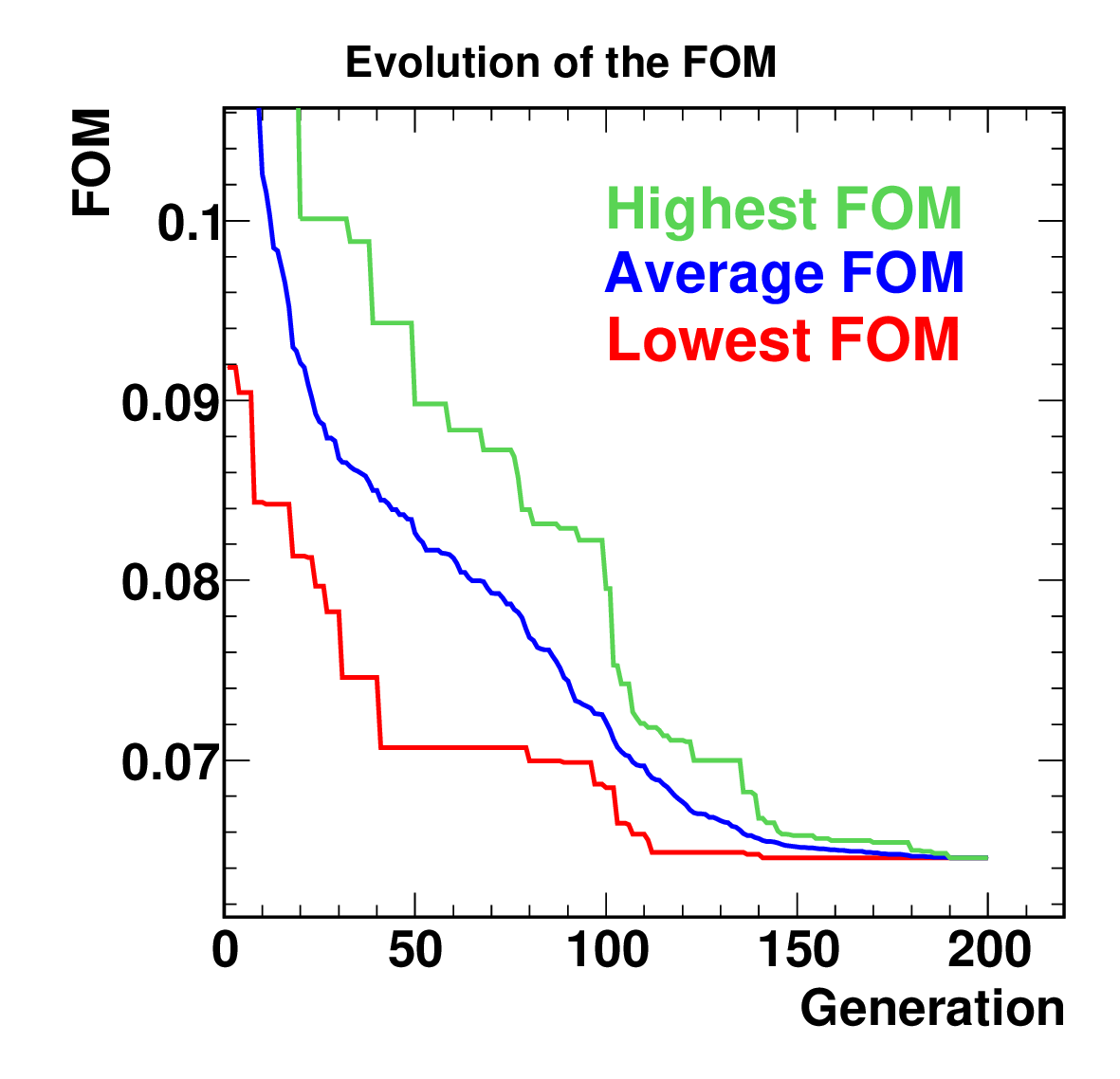} &
\includegraphics[width=0.45\linewidth]{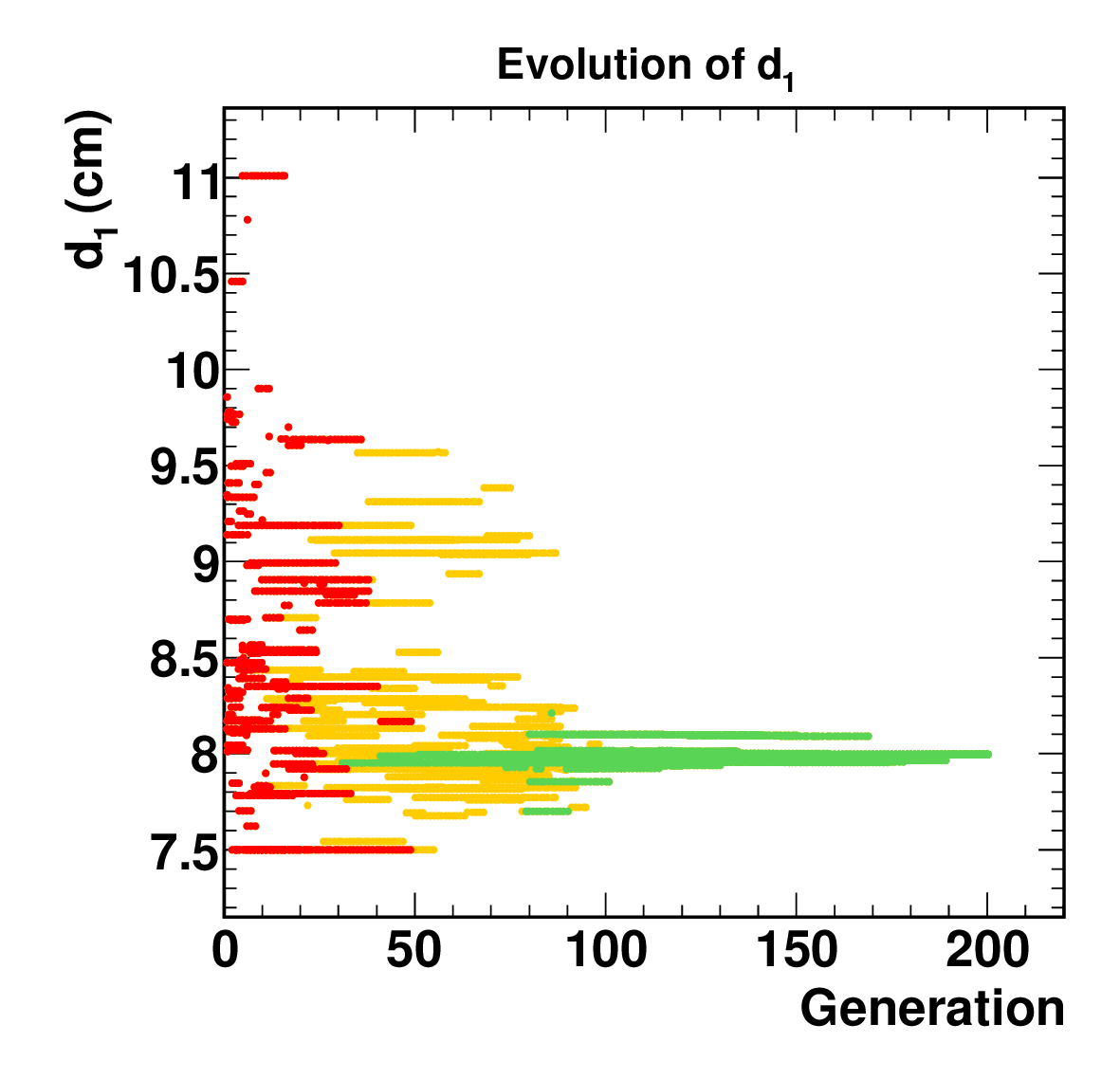} \\
\includegraphics[width=0.45\linewidth]{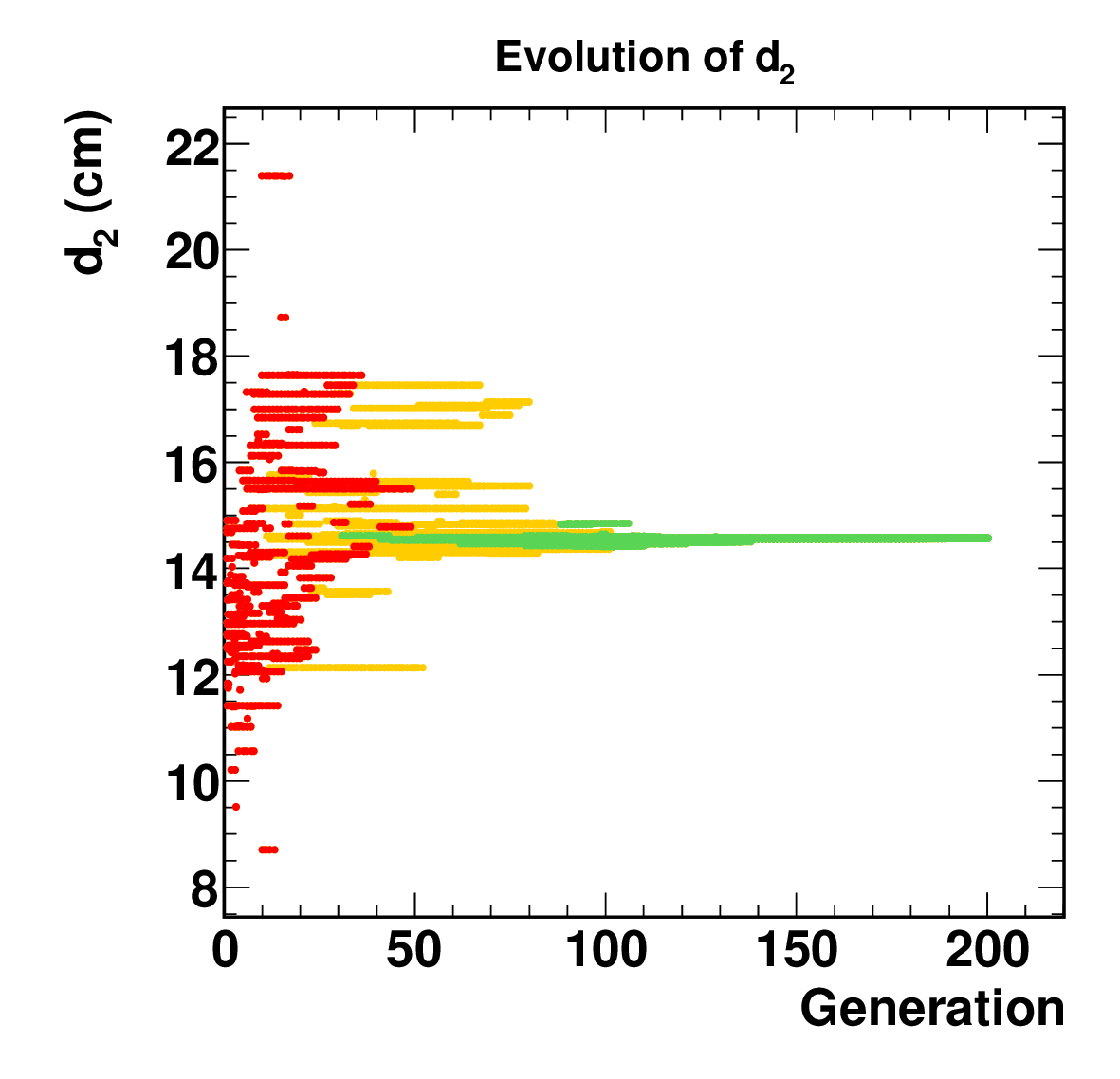} &
\includegraphics[width=0.45\linewidth]{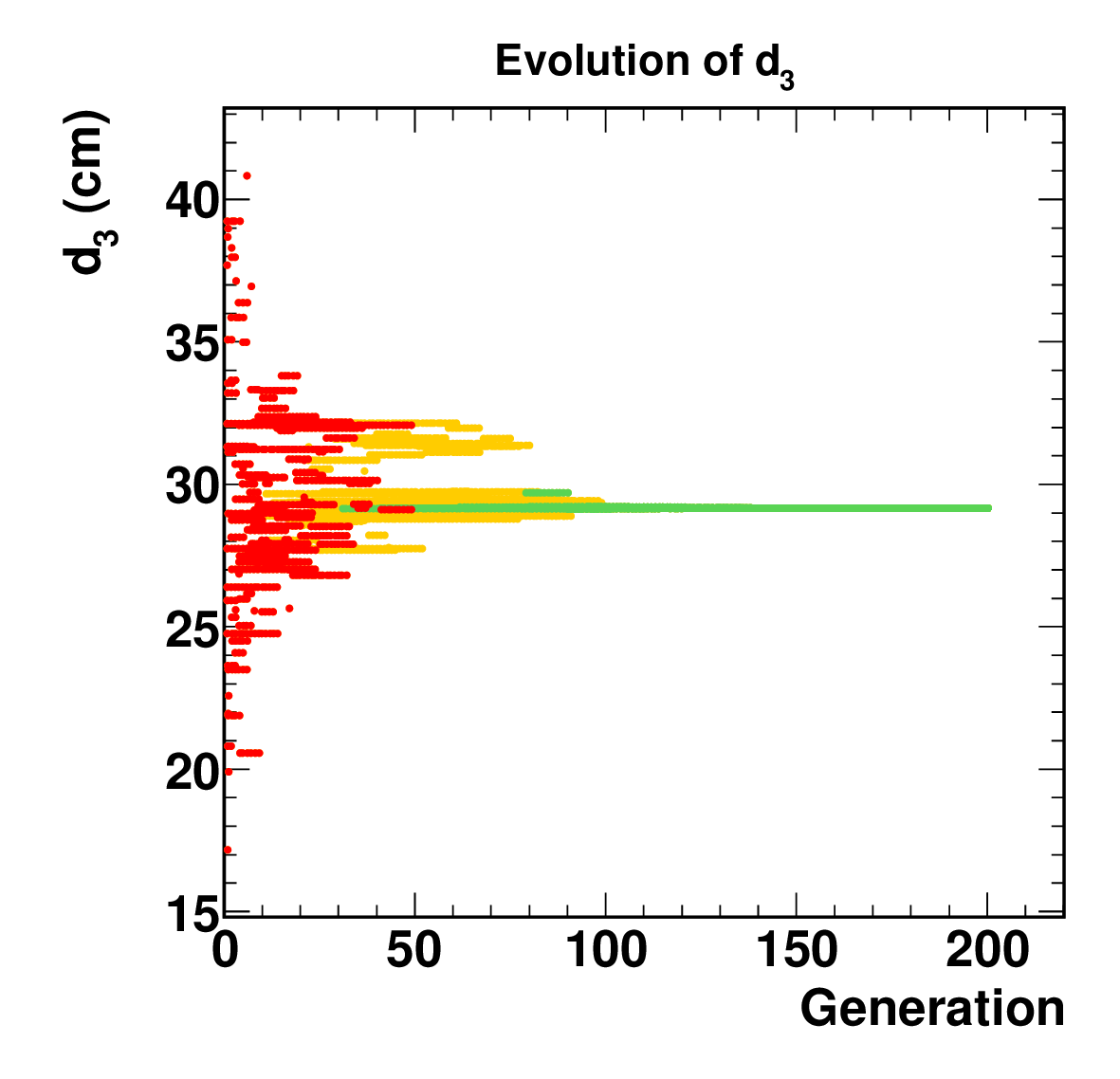} \\
\end{array}$
\end{center}
\caption{Example of fit performed with the differential evolution algorithm. In the top left panel we show the evolution of the figure of merit as a function of the generation (i.e. iteration). In the other three panels, the evolution also as a function of the generation, of the three parameters which are being fitted: $d_{1}$, $d_{2}$, and $d_{3}$. More details are given in the body text.}
\label{fig:FitExample01}
\end{figure*}

To minimize the figure of merit, we optimized the three parameters $d_{1}$, $d_{2}$, and $d_{3}$ using the differential evolution (DE) algorithm~\cite{Storn_1997,Das_2011,Das_2016}, following the methodology we already used in our previous work~\cite{Mendoza_2020}. For this optimization, we set the $\alpha$ value to 0.5 and the maximum neutron source energy to $E_{n,\max}=8$~MeV; note that while $E_{n,\max}$ depends on the specific measurement to be performed, its value is known beforehand. The DE algorithm is particularly well-suited for this task because evaluating the $\mathit{FOM}$ function at any point $(d_{1}, d_{2}, d_{3})$ requires a Monte Carlo simulation. Consequently, the resulting $\mathit{FOM}$ values have limited precision due to counting statistics, making a derivative-free minimization approach highly preferable. Furthermore, the DE algorithm is notably robust and scales efficiently in parallel computing environments.

The DE algorithm has a number of candidate solutions, referred to as \textit{agents}, which are iteratively improved. In each iteration (referred to as a \textit{generation}), new possible solutions, built from the agents, are tested. If these new candidate solutions improve upon the previous ones, the agents are updated.

For the fit, we used 30 agents (candidate solutions) that evolve over 200 iterations (generations). For each generation we have then evaluated 30 new values for the $FOM$, which involve making 30 new Monte Carlo calculations. This requires considerable computing power, for which we used the CIEMAT's Xula supercomputer. In each calculation, we simulated 10$^{7}$ source neutrons.

The result of the fitting process is presented in Fig.~\ref{fig:FitExample01}. There we present the values of the $FOM$, $d_{1}$, $d_{2}$, and $d_{3}$ as a function of the generation. Since we are using 30 agents, the figure shows, for each generation, 30 different values of $d_{1}$, $d_{2}$, and $d_{3}$, each of them corresponding to a different agent. In the case of the $FOM$, we show the largest value of each generation among the 30 $FOM$ values in green, the lowest value in red, and the average value in blue. Each of the points corresponding to the values of $d_{1}$, $d_{2}$, and $d_{3}$ appears in red if the value of the $FOM$ of the corresponding agent is larger than 0.090, in orange if it is between 0.075 and 0.090, and in green if it is smaller than 0.075. As can be observed, the values of $d_{1}$, $d_{2}$, and $d_{3}$ converge to the values reported in the previous section: $d_{1}=8.00$, $d_{2}=14.6$ and $d_{3}=29.2$ cm.

\section{Design of nSPIRAL} \label{sec:Design_nSPIRAL}

The fundamental purpose of long neutron counters is to measure neutron yields in nuclear reactions or $P_{n}$ values in radioactive decays. To measure neutron spectra, the usual practice is to perform time-of-flight measurements using other types of detectors~\cite{PerezDeRada_2025_v01,Jacobs_1983}. However, long neutron counters are also used to provide some information about neutron energy spectra. In this regard, the most common procedure has consisted of estimating the average neutron energy from the ratio of counts between the different rings~\cite{Reeder_1977,MontGeli_2026,Brandenburg_2022}.

Obtaining some information about neutron energies is possible because the response functions of the different rings have different shapes, as can be seen in Fig.~\ref{fig:EffExample01} (the curves labeled as R1, R2 and R3). The more counts in the outer ring with respect to the inner ring, the larger the energy of the neutrons being detected.

\begin{figure}[htb]
\begin{center}
\includegraphics[width=0.99\linewidth]{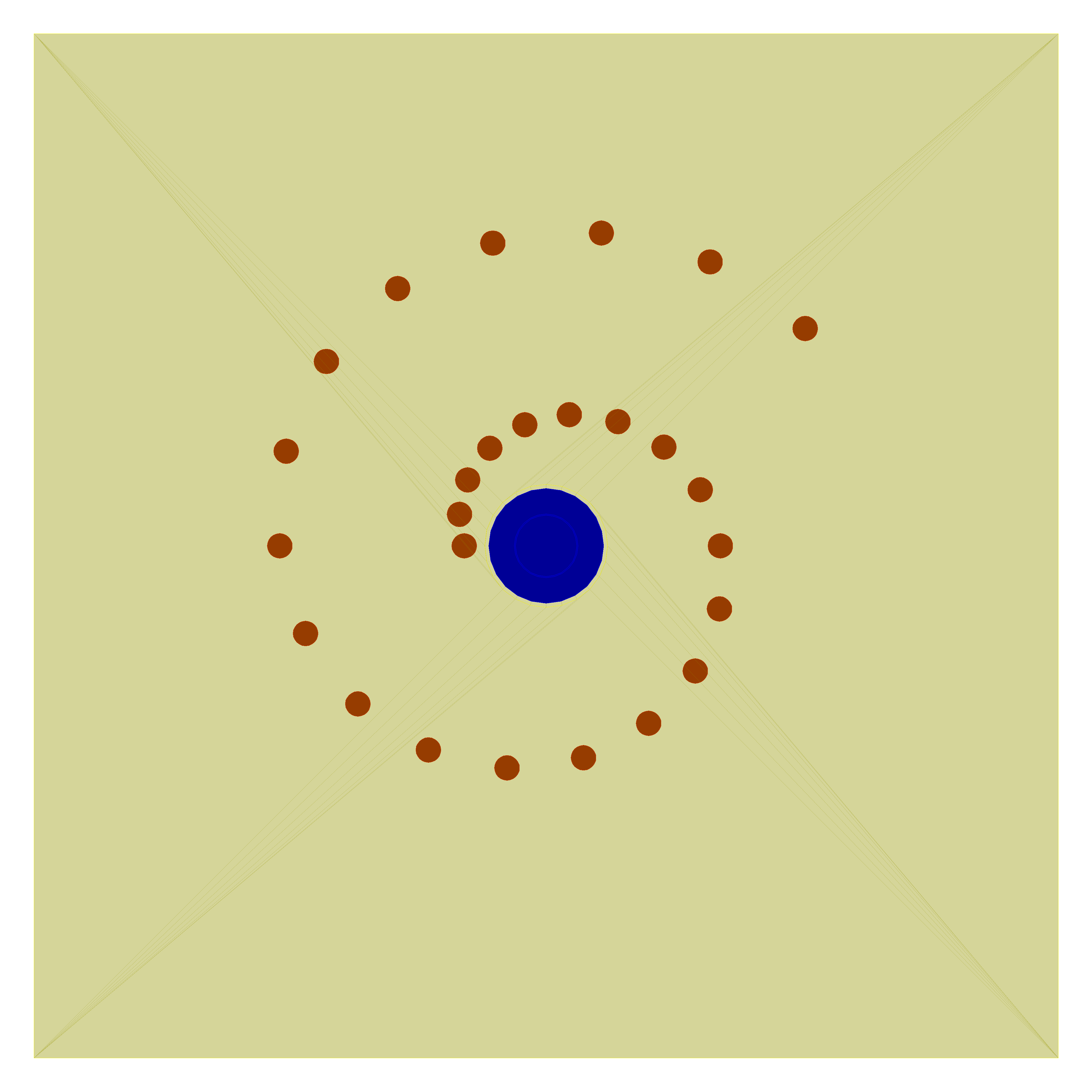} \\
\end{center}
\caption{Schematic view of a neutron long counter
with 26 $^{3}$He tubes arranged in a spiral. The distance from the $^{3}$He tubes to the center of the geometry is between 8 and 33~cm, in steps of 1~cm.}
\label{fig:Conf-Spiral}
\end{figure}

In principle, it is possible to obtain more detailed information about the neutron spectrum beyond just the average energy. To this end, the use of thermal neutron detectors combined with polyethylene moderators for neutron spectroscopy is widespread. The most prominent example is the Bonner sphere spectrometer~\cite{Bramblett_1960}, though numerous other configurations have been developed~\cite{Tanimura_2005,Stamati_2023,Lotfalizadeh_2017,Tursinah_2024,Vega-Carrillo_2026}. The underlying principle of these systems is that higher-energy neutrons penetrate deeper into the polyethylene matrix. The concept of using long counters as neutron spectrometers has been explored by some of the co-authors of this work to the specific case of the BELEN detector~\cite{Calvino_2023}; and more recently to the TETRA detector~\cite{Cantacuzene_2026}.

\begin{figure*}[t]
    \centering
    \begin{subfigure}[b]{0.46\textwidth}
        \centering
        \includegraphics[width=\textwidth]{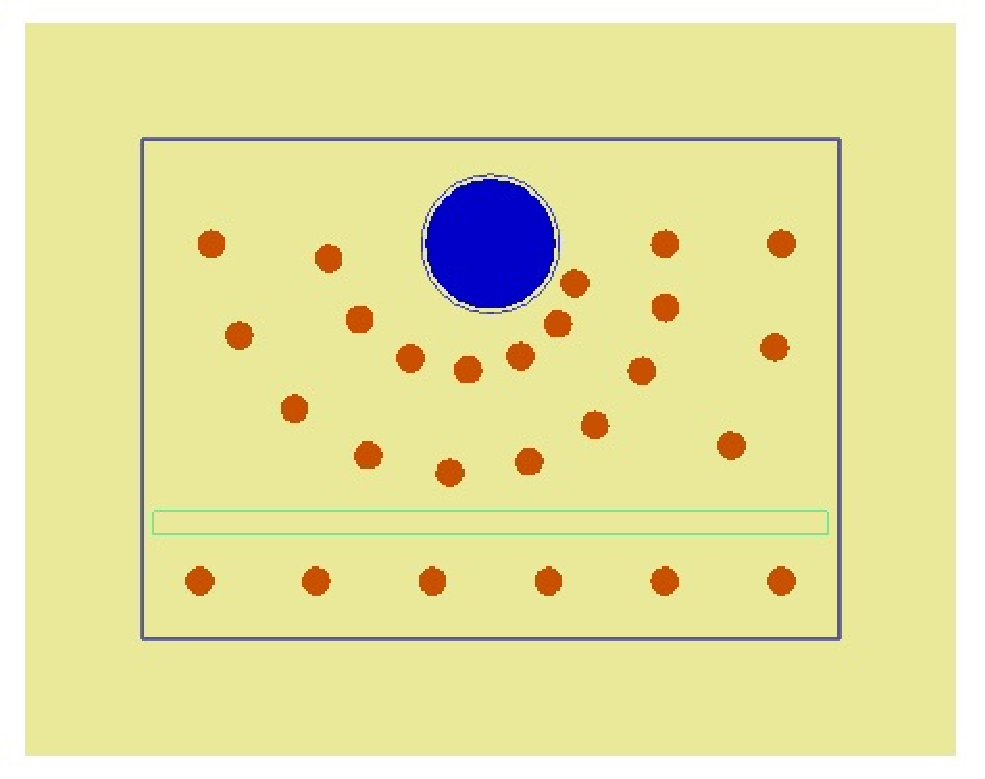} 
    \end{subfigure}
    \hfill 
    \begin{subfigure}[b]{0.53\textwidth}
        \centering
        \includegraphics[width=\textwidth]{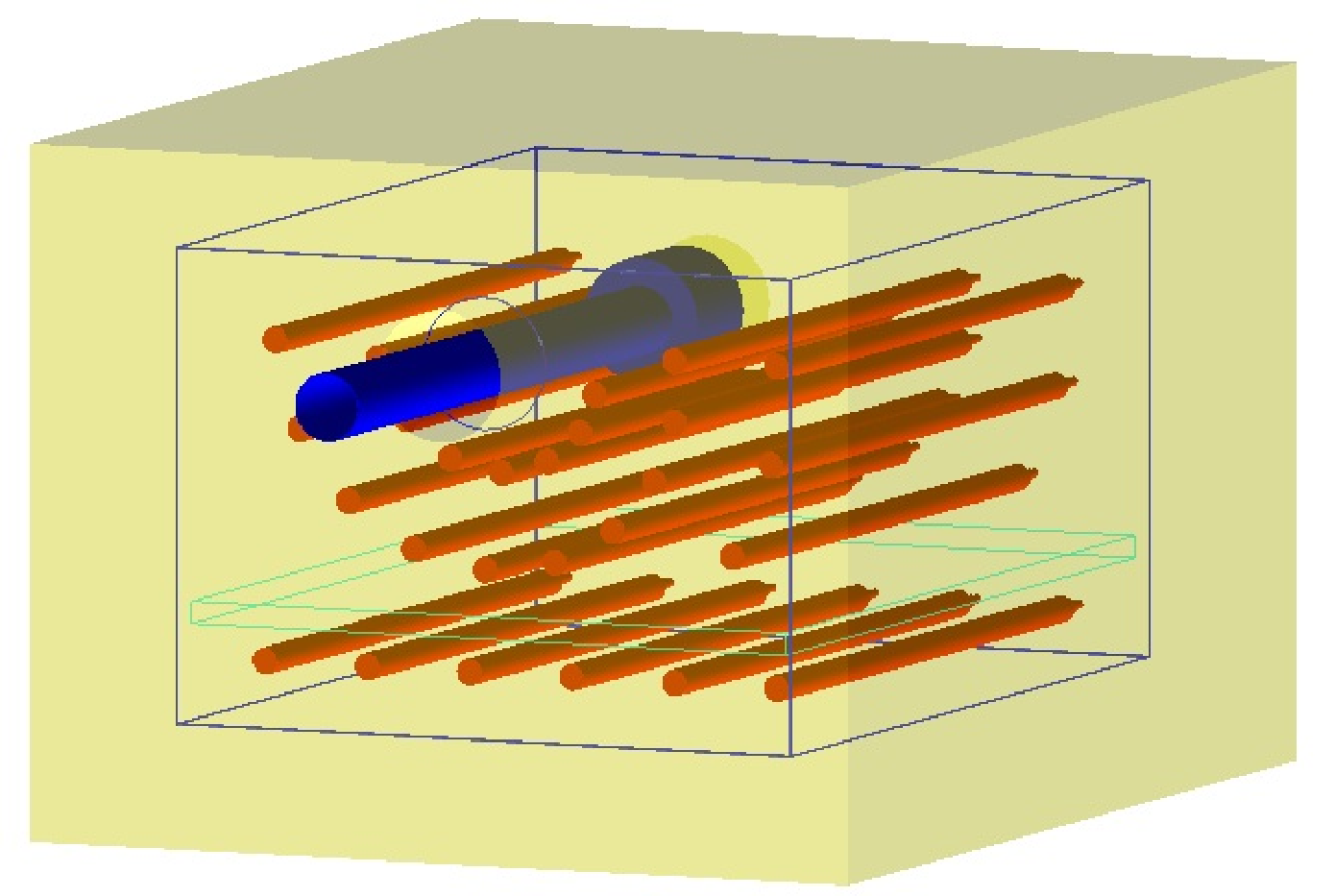} 
    \end{subfigure}
    \caption{Schematic view of nSPIRAL, from two different angles.}
    \label{fig:Conf-nSPIRAL}
\end{figure*}

To obtain more information about the spectrum of the neutrons being measured, it seems convenient to place the $^{3}$He tubes not in rings but at many different distances from the neutron source, with the idea of having a larger amount of different response functions. Using the weighting technique allows for great flexibility in the design of the detection device, since, once the geometry is defined, the weights can be adjusted to achieve a flat detection efficiency.

One possibility is the one shown in Fig.~\ref{fig:Conf-Spiral}, where the 26 $^{3}$He are arranged in a spiral instead of rings. If we start from the detector closer to the center, each additional detector is located 1 cm further from the center than the previous one and rotated by 20 degrees.

Following this idea and after testing several possibilities, we have converged to the design shown in Fig.~\ref{fig:Conf-nSPIRAL}. The spiral is still present here, but it is located entirely below the neutron source, rather than surrounding it. This aims to reduce the size of the detection device, and therefore make it easier to handle. The 26 $^{3}$He tubes are all located inside a 60x43x80 cm$^{3}$ parallelepiped made of polyethylene, surrounded by 0.5 mm thick Cd layers. After the Cd, we put down a 10 cm thick polyethylene shielding, so the outer dimensions of the polyethylene block are 80x63x100 cm$^{3}$. Regarding the position of the $^{3}$He tubes, there are 20 of them located in three sub-spirals at distances from the center of between 8 and 27 cm, in steps of 1 cm. Below them, there are 6 additional tubes, located out of the spiral structure to be able to place a Pb neutron converter of 2 cm thick between them and the neutron source. This Pb layer is shown in Fig.~\ref{fig:Conf-nSPIRAL} as a parallelepiped with cyan lines.

If the neutrons from the source have energies up to 8 or 10 MeV at most, this Pb converter is not necessary. However, if the source emits neutrons of higher energies, the Pb will allow for a flatter detection efficiency. This is because the cross section of $^{1}$H becomes smaller as the neutron energy increases, making the polyethylene less effective at moderating higher energy neutrons. Placing a Pb converter, which is a standard procedure~\cite{Wiegel_2002,Gomez-Ros_2012}, causes (n,xn) reactions to occur, which reduce the energy of these high energy neutrons. The idea is for the detector device to operate in two configurations, with or without a Pb converter, depending on the maximum energy of the neutrons to be measured in the experiment. We will refer to the configurations without and with Pb as nSPIRAL and nSPIRAL-Pb, respectively.

Regarding the shielding, composed of cadmium and 10 cm of polyethylene, it serves a dual purpose. The first is to protect the detector from externally incident neutrons. The second is to prevent neutrons generated within the detection device from escaping, re-entering, and triggering a count in the $^{3}$He tubes. Minimizing this second effect is crucial; otherwise, the detector response functions $\varepsilon_{i,j}$, derived via Monte Carlo simulations, would become dependent on the surrounding material, a situation we wish to avoid at all costs. The shielding thickness was designed to satisfy this second condition. To this end, simulations were conducted with the detector surrounded by a graphite reflector, testing various shielding thicknesses. The results showed that 10 cm of polyethylene provides sufficient shielding to ensure the $^{3}$He tube response functions remain minimally affected by the graphite reflector.

\begin{figure}[htb]
\begin{center}
\includegraphics[width=0.99\linewidth]{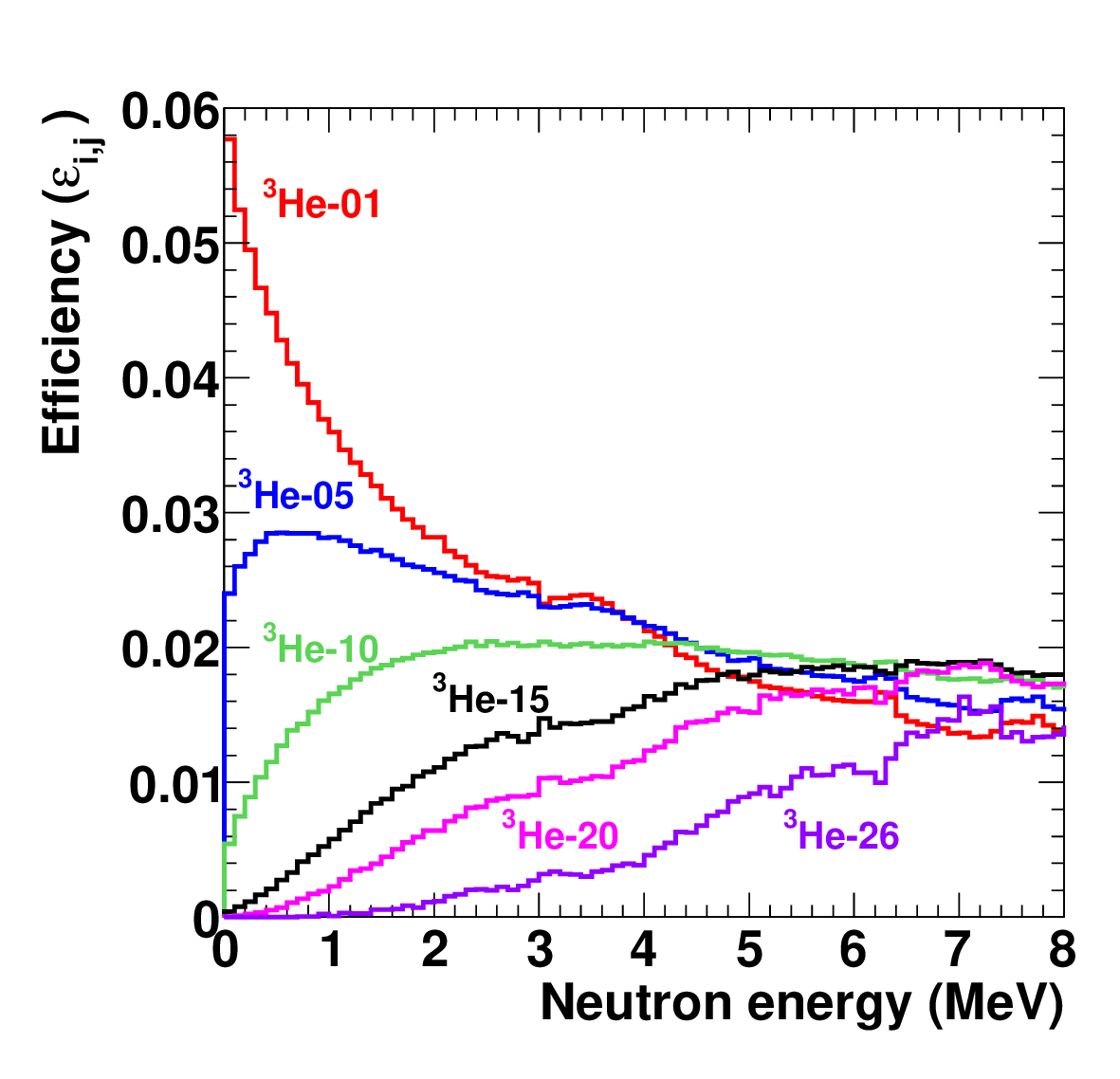} \\
\end{center}
\caption{Response functions of six different $^{3}$He tubes of the nSPIRAL detector (without Pb converter). They are labeled as $^{3}$He-01, $^{3}$He-05 ... meaning that if we order the $^{3}$He tubes by distance to the center, $^{3}$He-01 is the tube closer to the neutron source, $^{3}$He-05 is the fifth closest, etc. $^{3}$He-26 is the detector in the bottom-right corner in Fig.~\ref{fig:Conf-nSPIRAL}.}
\label{fig:ResponseFunctions01}
\end{figure}

We show the response functions of some of the $^{3}$He tubes in Fig.~\ref{fig:ResponseFunctions01}. As can be seen, the response functions are different. This is what makes neutron spectroscopy possible.

\begin{figure*}[t]
    \centering
    \begin{subfigure}[b]{0.32\textwidth}
        \centering
        \includegraphics[width=\textwidth]{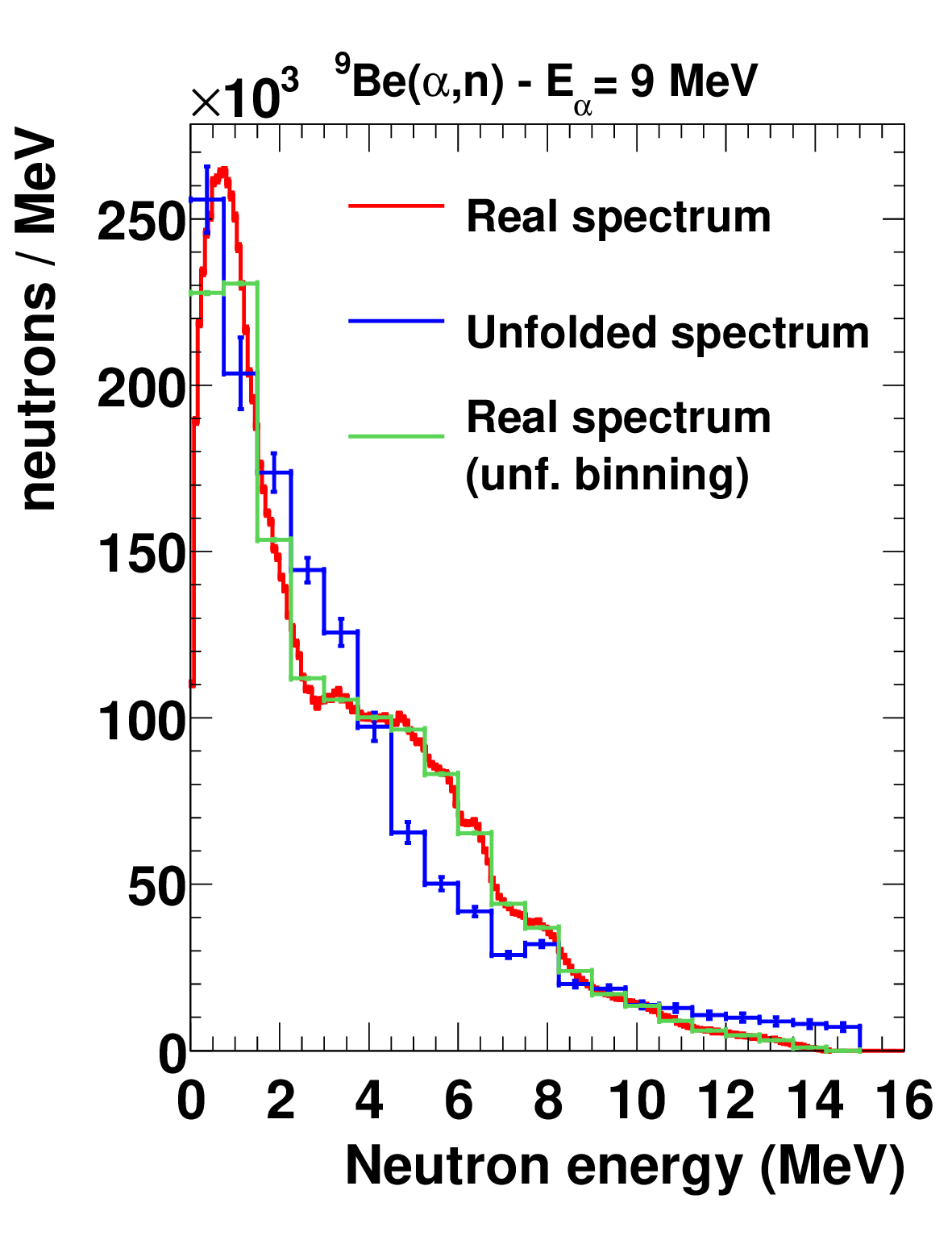} 
    \end{subfigure}
    \hfill 
    \begin{subfigure}[b]{0.32\textwidth}
        \centering
        \includegraphics[width=\textwidth]{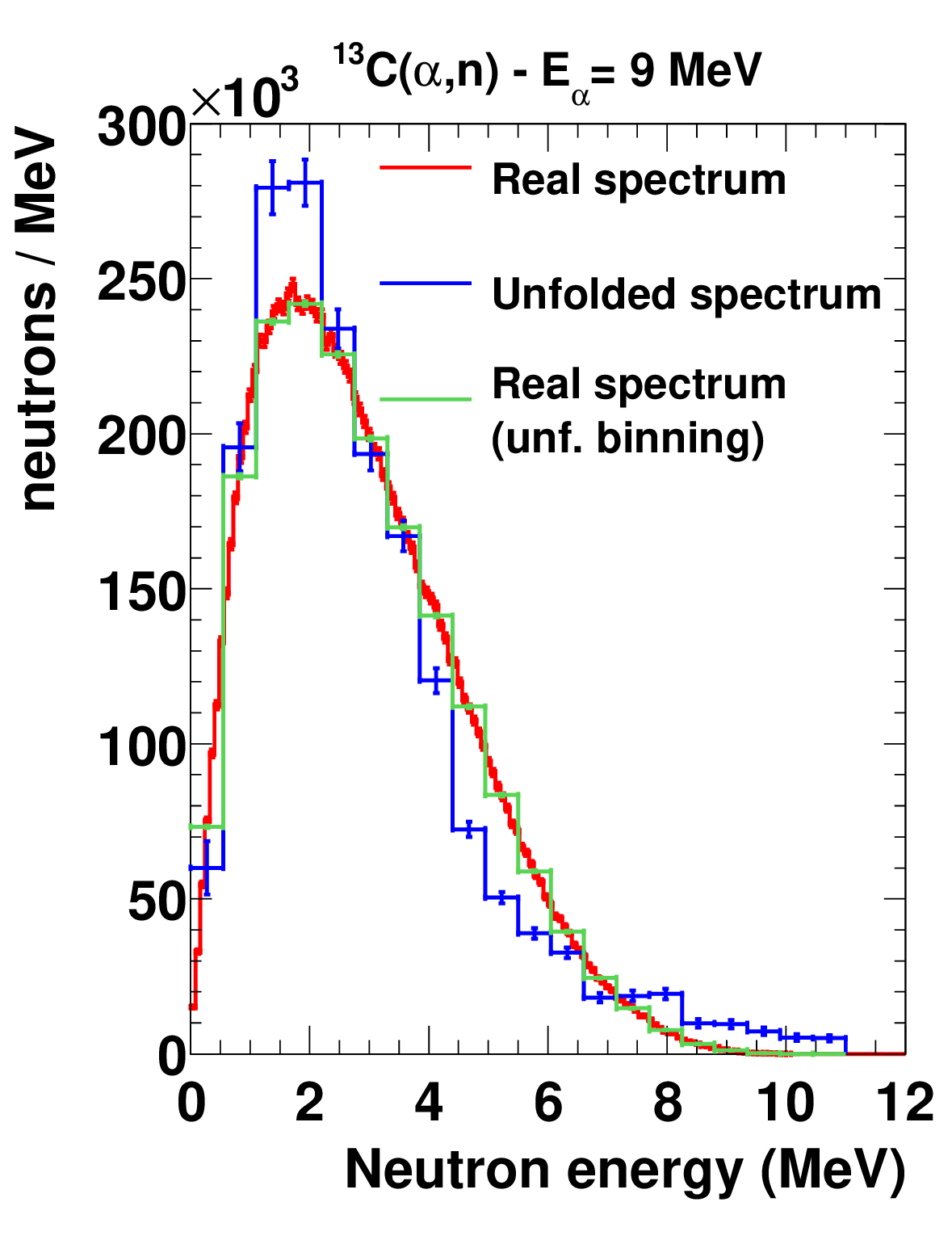} 
    \end{subfigure}
    \begin{subfigure}[b]{0.32\textwidth}
        \centering
        \includegraphics[width=\textwidth]{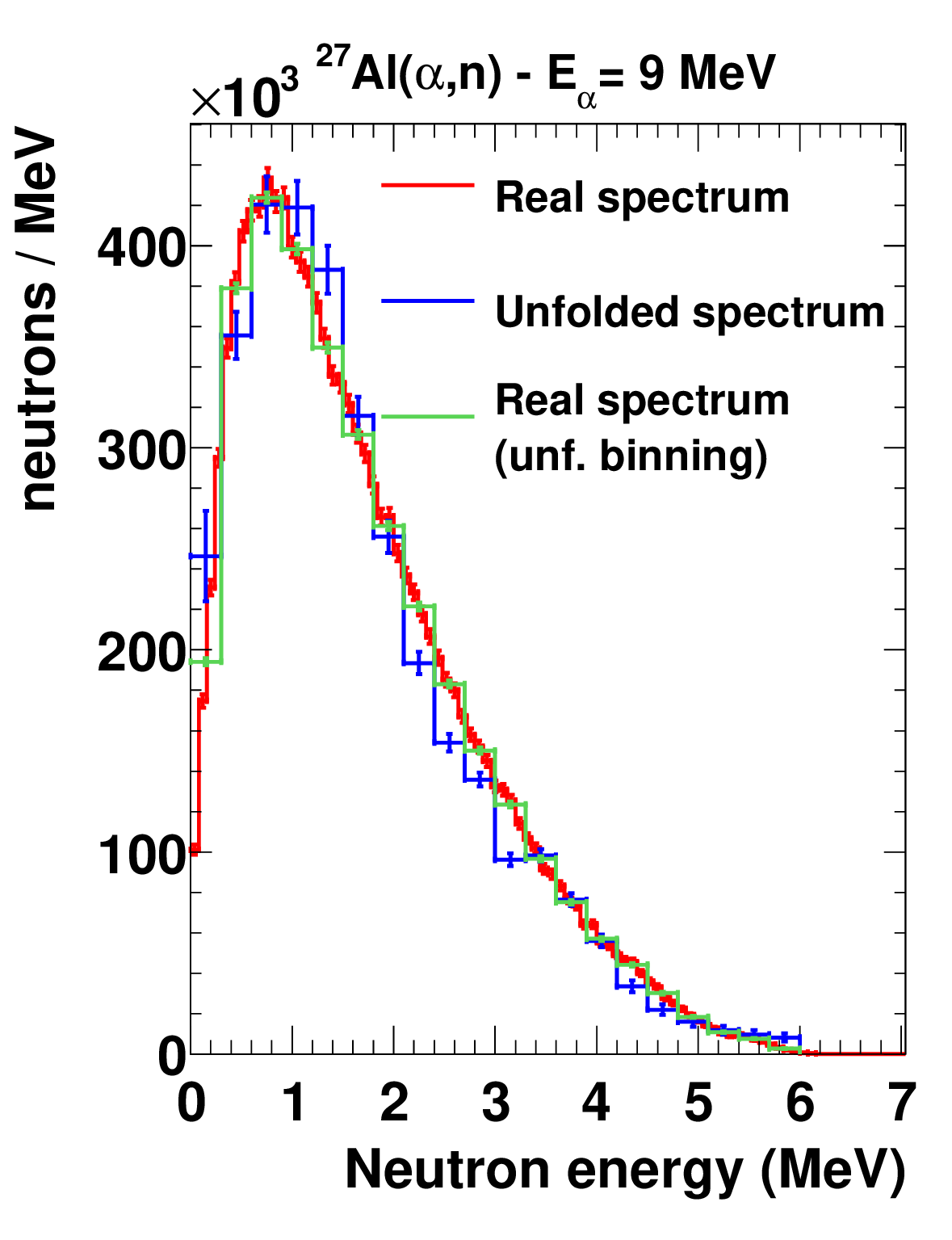} 
    \end{subfigure}
    \caption{Examples of unfolding performed with RooUnfold. Shown in red (\textit{Real spectrum}) are the true simulated spectra, corresponding to neutrons generated when a 9 MeV $\alpha$-particle beam impinges into thick $^{9}$Be (left) $^{13}$C (middle) and $^{27}$Al (right) targets. The unfolded spectra are shown in blue (\textit{Unfolded spectrum}), while the green (\textit{Real spectrum (unf. binning)}) curves represent the true simulated spectra expressed in bins of the unfolded spectra. In all three cases the unfolding started with a uniform initial spectrum.}
    \label{fig:DeconvolutionExample01}
\end{figure*}

In the following section, we present the performance of this detector when operating as a neutron counter. Let us now examine its spectroscopic capabilities. To this end, we simulated the detection of neutrons originating from ($\alpha$,n) reactions. We employed SaG4n (version 1.6)~\cite{Mendoza_2020_v02,SaG4n_website} to determine the neutrons emitted from thick samples when subjected to an incident $\alpha$-particle beam of a specific energy. The list-mode output from SaG4n, containing the position, direction, and energy of the produced neutrons, was then used as the input source for the Geant4 simulations ($10^{6}$ source neutrons per simulation). The result of these simulations is a specific number of counts in each of the 26 $^{3}$He tubes.

We have performed simulations of thick monoisotopic samples located at the center of nSPIRAL using most of the nuclei included in the JENDL/AN-2005 library~\cite{Murata_2006,JENDL-AN-2005_website}: $^{6,7}$Li, $^{9}$Be; $^{10,11}$B, $^{12,13}$C, $^{14,15}$N, $^{17,18}$O, $^{19}$F, $^{23}$Na, $^{27}$Al and $^{28,29,30}$Si. In SaG4n, we have used JENDL/AN-2005 to model the ($\alpha$,n) reactions. The response matrix of the detection device, $\varepsilon_{i,j}$, has been calculated using the same Geant4 geometry, but as usual with the neutrons emitted isotropically from the center of nSPIRAL. $\varepsilon_{i,j}$ relates the number of counts in each tube, $C_{i}$, to the number of neutrons emitted in each energy range, $N_{S,j}\simeq N_{S}\cdot p_{S,j}$:

\begin{equation} \label{eq:RMatrix}
C_{i}\simeq \sum_{j}\varepsilon_{i,j}\cdot N_{S,j}
\end{equation}

Let us recall that $N_{S}$ is the number of neutrons emitted by the source $S$ in a certain period of time, and that in Eq.~\ref{eq:RMatrix} the $i$-index refers to the $^{3}$He tube number and the $j$-index to the energy interval. The symbols $\simeq$ appear in the expressions above instead of $=$ due to the statistical fluctuations. 

The task now is to derive $N_{S,j}$ (the energy spectrum) from $\varepsilon_{i,j}$ (the response matrix) and $C_{i}$ (the number of counts in the $^{3}$He tubes). This is what is known as unfolding the neutron spectrum, and there are numerous methods for carrying it out~\cite{Reginatto_2010}. We have used RooUnfold~\cite{Brenner_2020,RooUnfold_website}, a code which can be used with ROOT~\cite{root_1997}. To perform the unfolding, we used the \textit{RooUnfoldBayes} algorithm, which employs the iterative Bayesian method described in~\cite{Agostini_1995}, with which we have experience~\cite{PerezDeRada_2025_v02}. As an initial solution for the iterative method, we have in all cases adopted a uniform energy spectrum ranging from 0 to some value a bit larger than the maximum possible neutron energy.

Overall, the unfolding results demonstrate satisfactory performance. Fig.~\ref{fig:DeconvolutionExample01} illustrates three representative examples corresponding to the $^{9}$Be, $^{13}$C, and $^{27}$Al targets under 9~MeV incident $\alpha$-particles. In all cases, the true simulated energy spectrum (red) is presented alongside the unfolded spectrum (blue). For a direct comparison, the simulated spectrum is also shown using the same binning configuration as the unfolding procedure (green). The remaining cases tested exhibit comparable behavior to these three examples. Notably, the response matrix was calculated assuming an isotropic neutron source distribution, whereas the neutrons emitted in the actual ($\alpha$,n) reactions are anisotropic. Additional simulations performed with isotropic emission, maintaining the same reaction-specific energy distributions, generally yielded slight improvements in the unfolding accuracy.

\section{Performance of nSPIRAL} \label{sec:Performance_nSPIRAL}

In this work we have presented four distinct detection device configurations: the optimized three-ring design; the spiral design; and the nSPIRAL design, without and with lead. We will refer to these four configurations, summarized in Table~\ref{tab:Conf-description}, respectively, as Conf-3R, Conf-Spiral, nSPIRAL and nSPIRAL-Pb. Next, we will compare the performance of these four configurations.

\begin{table}[ht]
\centering
\begin{tabularx}{\columnwidth}{lX}
\toprule
Configuration & Description \\
\midrule
Conf-3R & Detectors located in three rings at optimized distances from the center, as shown in Fig.~\ref{fig:Conf-3R_02}. \\
Conf-Spiral & Detectors arranged in a spiral, as shown in Fig.~\ref{fig:Conf-Spiral}. \\
nSPIRAL & Configuration shown in Fig.~\ref{fig:Conf-nSPIRAL}, without Pb neutron converter. \\
nSPIRAL-Pb & Configuration shown in Fig.~\ref{fig:Conf-nSPIRAL}, with Pb neutron converter. \\
\bottomrule
\end{tabularx}
\caption{Description of four different configurations of neutron long counters discussed in this work.}
\label{tab:Conf-description}
\end{table}

\begin{figure}[htb]
\begin{center}
\includegraphics[width=0.99\linewidth]{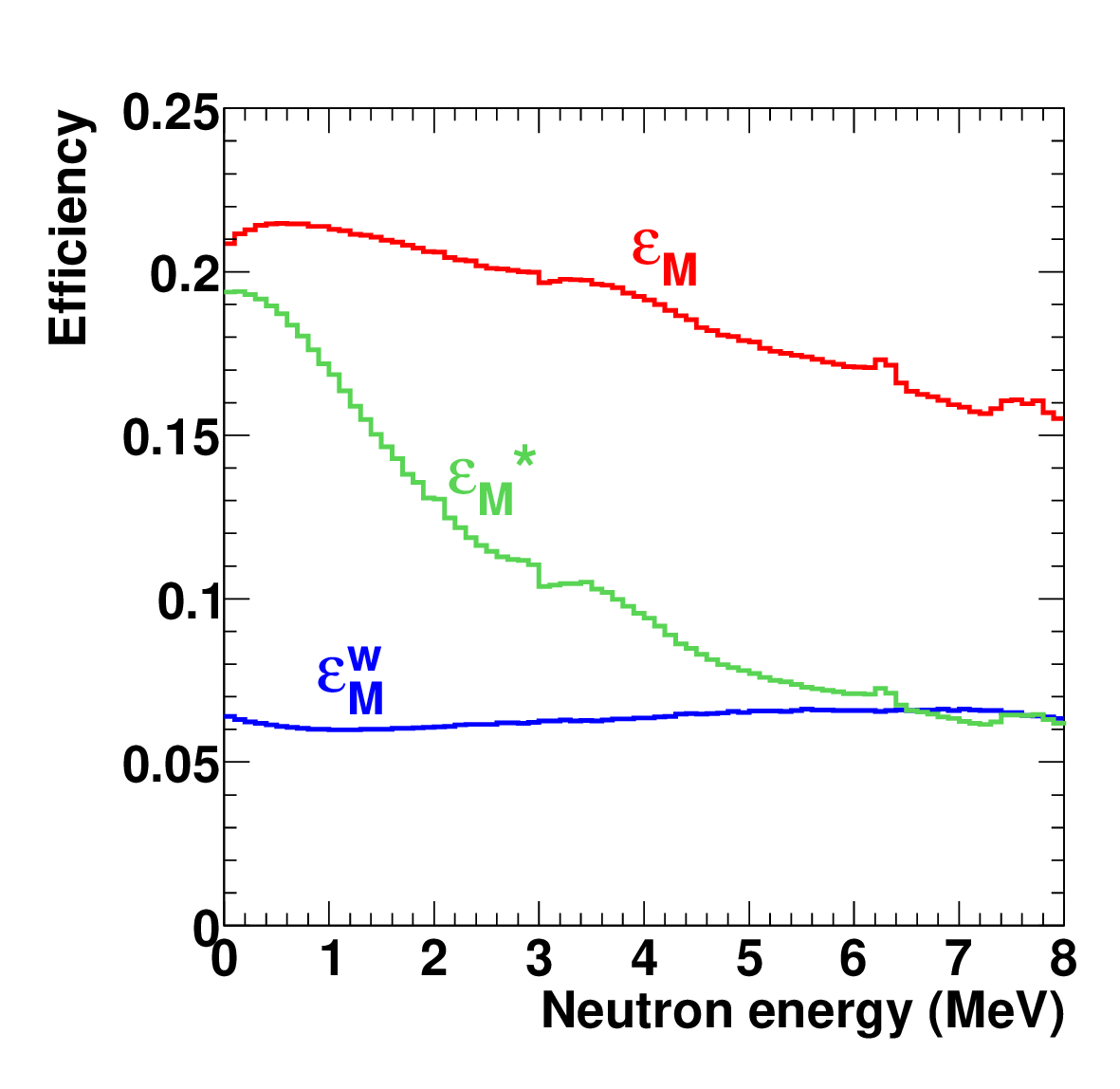} \\
\end{center}
\caption{Different efficiencies of the nSPIRAL detector, equivalent to the ones shown in Fig.~\ref{fig:EffExample01} for Conf-3R. The weights have been fitted to achieve a flat efficiency between 0 and 8 MeV.}
\label{fig:Eff-Conf862}
\end{figure}

The various efficiencies defined in this work as a function of neutron energy are shown in Fig.~\ref{fig:Eff-Conf862} for nSPIRAL. This figure is analogous to Fig.~\ref{fig:EffExample01}, which was obtained for the Conf-3R configuration. Notably, nSPIRAL exhibits an efficiency $\varepsilon^{*}_{M}$ (the one that determines how much statistics we need for the measurement) that is $\sim$30\% lower than that of Conf-3R; however, its $\varepsilon^{w}_{M}$ profile appears to be remarkably flat.

The flatness factors $F_{1}$ and $F_{2}$ for the 4 configurations, along with the corresponding $\varepsilon^{*}_{S}$ values, are shown in Table~\ref{tab:Conf-Comp01} for different energy ranges provided by $E_{n,max}$. To compute $\varepsilon^{*}_{S}$ we have used a uniform distribution from 0 to $E_{n,max}$.

\begin{table}[ht]
\centering
\begin{tabular}{lccc}
\toprule
Configuration & $F_{1}$ & $F_{2}$ & $\varepsilon^{*}_{S}$ \\
\midrule
\multicolumn{4}{c}{$E_{n,\max} = 5$ MeV} \\
Conf-3R      & 1.01 & $7.5\cdot10^{-6}$  & 18.5\% \\
Conf-Spiral  & 1.01 & $1.4\cdot10^{-6}$  & 11.6\% \\
nSPIRAL        & 1.01 & $2.2\cdot10^{-6}$  & 12.8\% \\
nSPIRAL-Pb     & 1.01 & $2.9\cdot10^{-6}$  & 12.6\% \\
\midrule
\multicolumn{4}{c}{$E_{n,\max} = 8$ MeV} \\
Conf-3R      & 1.06 & $1.1\cdot10^{-4}$  & 14.9\% \\
Conf-Spiral  & 1.06 & $1.5\cdot10^{-4}$  & 12.0\% \\
nSPIRAL        & 1.06 & $1.0\cdot10^{-4}$  & 10.4\% \\
nSPIRAL-Pb     & 1.06 & $1.1\cdot10^{-4}$  & 10.4\% \\
\midrule
\multicolumn{4}{c}{$E_{n,\max} = 15$ MeV} \\
Conf-3R      & 1.14 & $1.4\cdot10^{-3}$  & 9.2\% \\
Conf-Spiral  & 1.18 & $2.4\cdot10^{-3}$  & 8.7\% \\
nSPIRAL        & 1.15 & $1.8\cdot10^{-3}$  & 6.5\% \\
nSPIRAL-Pb     & 1.10 & $4.3\cdot10^{-4}$  & 5.1\% \\
\bottomrule
\end{tabular}
\caption{Flatness factors $F_{1}$ and $F_{2}$ together with $\varepsilon^{*}_{S}$, for different long counter configurations and $E_{n,max}$. The weights for each long counter have been re-fitted for each $E_{n,max}$ value.}
\label{tab:Conf-Comp01}
\end{table}

\begin{figure}[!htb]
\begin{center}
\includegraphics[width=0.99\linewidth]{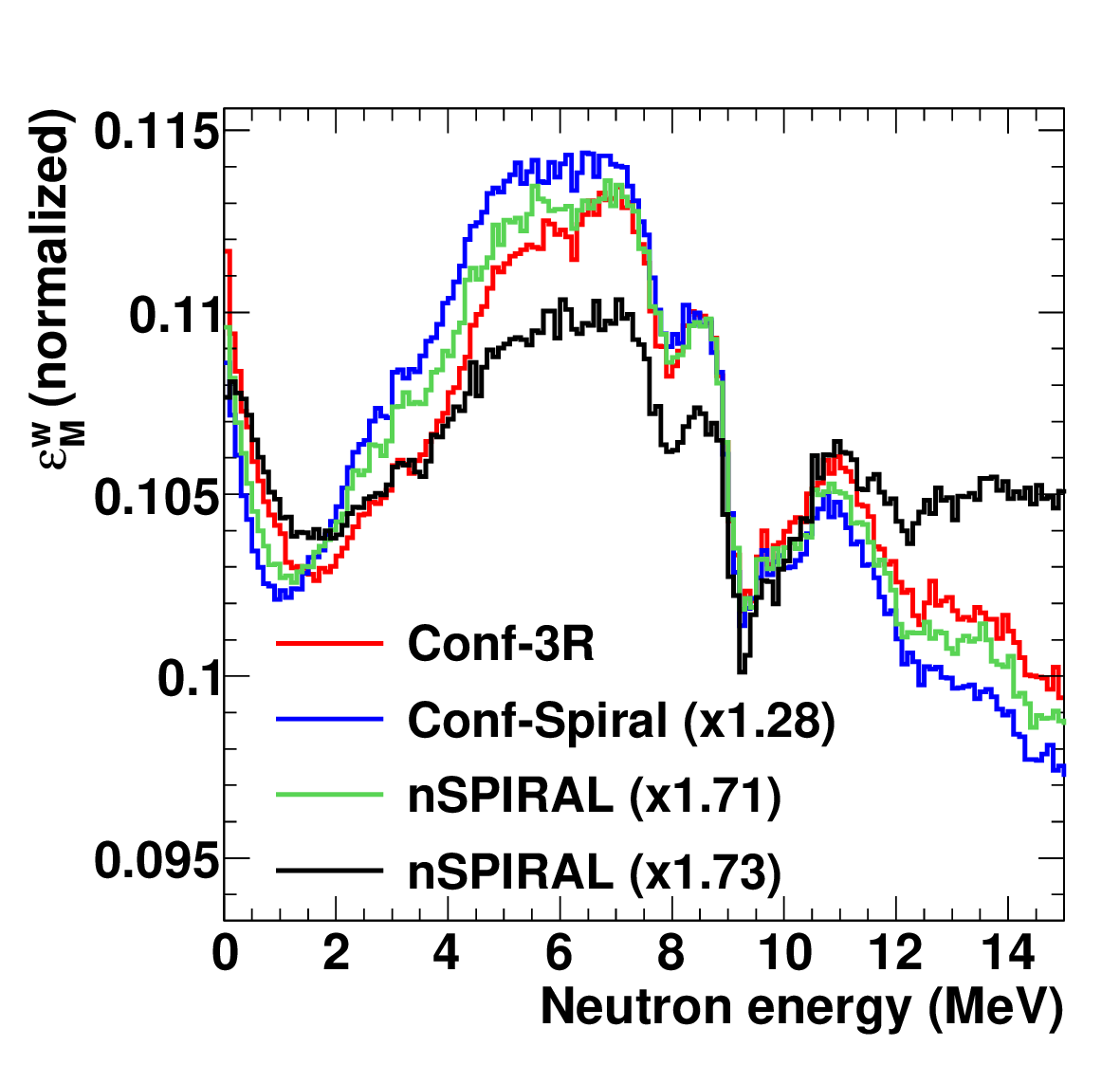} \\
\end{center}
\caption{$\varepsilon^{w}_{M}$ as a function of neutron energy for the four detection device configurations.}
\label{fig:Flatness4Conf}
\end{figure}

The $F_{1}$ values in Table~\ref{tab:Conf-Comp01} can be directly compared with those provided in Table 3 of~\cite{MontGeli_2026} (there $F_{1}$ is denoted as $F$) for several neutron long counters. Thanks to the weighting technique, the four detector configurations presented in this work achieve an efficiency flatness far superior to that reported for the other detectors. Regarding the differences between the four configurations, the most notable points are that: (i) Conf-3R has a larger efficiency ($\varepsilon^{*}_{S}$) in all cases; (ii) the performance of Conf-Spiral and nSPIRAL is similar, which indicates that making the detector more compact (nSPIRAL) does not result in a significant loss of performance; and (iii) the flatness of the efficiency is similar in all cases, with the exception of nSPIRAL-Pb, which performs better for $E_{n,max}=15$ MeV. We also show this last point in Fig.~\ref{fig:Flatness4Conf}, which presents $\varepsilon^{w}_{M}$ as a function of neutron energy for the four configurations, appropriately normalized to facilitate comparison among them.

Regarding the weights obtained with nSPIRAL, we show as an example those obtained when fitted in the range of 0 to $E_{n,max}=8$ MeV. If we order the $^{3}$He tubes by proximity to the neutron source, as in Fig.~\ref{fig:ResponseFunctions01}, the values obtained are: $w_{i}=0.30-0.35$ for the 6 $^{3}$He tubes closer to the source, i.e. for $i=1-6$, $w_{i}=0.20-0.30$ for $i=7-14$, $w_{i}=0.40-1.00$ for $i=15-20$, and $w_{i}=1.50-5.00$ for $i=21-26$. Regarding detection efficiencies, for a $^{252}$Cf neutron source, they range from $\sim 3.1$\% to $\sim 6.3\cdot10^{-3}$\%; and for a uniform distribution between 0 and 8 MeV, they range from $\sim 2.4$\% to $\sim 1.9\cdot10^{-2}$\%.

In Section~\ref{sec:Efficiency_Flatness} we defined $p_{S}(\vec{\lambda})$ as the probability distribution which defines the neutron source. There $\vec{\lambda}$ represents the source parameters unknown a priori. In the most general case $\vec{\lambda}=\{E_{n},\theta,\phi,\vec{x}\}$, but in all this work we have worked with $\vec{\lambda}=\{E_{n}\}$, i.e. that the only relevant source parameter to be considered is the neutron energy.

\begin{figure*}[htb]
\begin{center}
\includegraphics[width=0.99\linewidth]{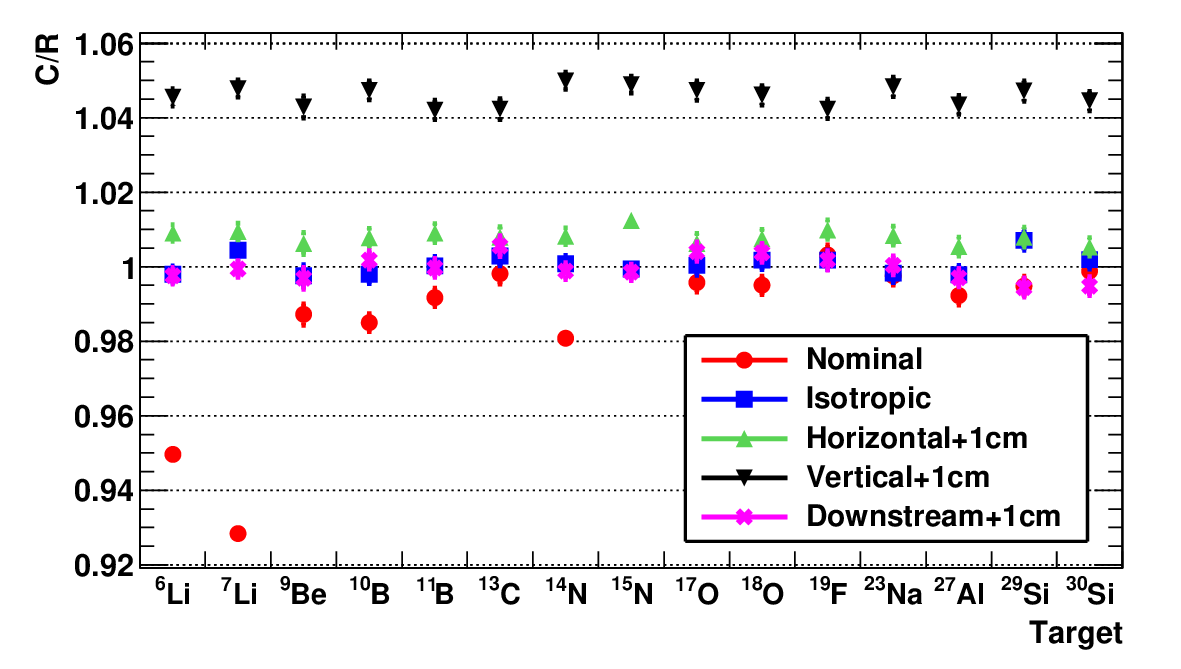} \\
\end{center}
\caption{Ratios between the number of source neutrons calculated with nSPIRAL (C) from the number of counts in the $^{3}$He tubes (using Eq.~\ref{eq:Estimator2}) and the real (R=10$^{6}$) number of source neutrons in simulations of ($\alpha$,n) reactions with various thick samples, in all cases using 9 MeV incident $\alpha$-particles. The results are presented for five different scenarios: the nominal case, with the source neutrons emitted from the center of nSPIRAL and using the energy-angle distributions from SaG4n; \textit{Isotropic}, which is identical to the nominal case but with neutrons emitted isotropically; and three additional cases, where the neutrons have been emitted isotropically, and with the source position shifted by 1 cm along each of the Cartesian axes. The point corresponding to the nominal case for $^{15}$N falls outside the visible portion of the graph, and is at C/R$\simeq 0.88$.}
\label{fig:AngPosDepend01}
\end{figure*}

This has also been the primary concern in the design of other neutron long counters~\cite{MontGeli_2026,Tarifeno-Saldivia_2017,Tolosa-Delgado_2019,Pereira_2010,Agramunt_2016,Mathieu_2012,Csedreki_2021,Testov_2016,Brandenburg_2022,Macklin_1957}. However, if the uncertainty in the angular distribution or the position of the emitted neutrons in our experiment is significant, it is preferable that our detection device not be highly dependent on these variables. The dependence of neutron long counters efficiency on angular distribution is addressed, for example, in~\cite{Li_2022,Csedreki_2021,Brandenburg_2022}. In these three cases, there is concern regarding the angular dependence of the neutrons emitted in the $^{13}$C($\alpha$,n) reaction. Indeed, it is concluded that corrections must be made to account for this angular dependence~\cite{Brandenburg_2023,Csedreki_2024}. The dependence on the position is addressed, for example, in~\cite{MontGeli_2026}.

Fig.~\ref{fig:AngPosDepend01} shows the impact of the angular distributions and the initial position of the neutrons in nSPIRAL. It presents the results of simulating the detection of neutrons produced by a 9 MeV $\alpha$-particle beam in various thick monoisotopic samples. The isotopes are those of the JENDL/AN-2005 library, and the neutrons were obtained from SaG4n when using JENDL/AN-2005. 10$^{6}$ source neutrons were simulated in each case, and Eq.~\ref{eq:Estimator2} was used to estimate, from the number of counts in the $^{3}$He tubes, the number of source neutrons, using the weights of nSPIRAL corresponding to $E_{n,max}=8$ MeV, which gives $\varepsilon^{w}_{S}=0.0669$. It should be noted that we also obtained the number of source neutrons from unfolding analyses, as these provide not only the neutron spectrum but also the total number of neutrons. We obtained values very similar to those shown below.

The blue squares labeled as \textit{Isotropic} show the case where neutrons have been emitted from the center and with an isotropic angular distribution, with the energy distribution provided by SaG4n. In this case deviations from the real number of source neutrons are smaller than 1\% in all cases (including $^{9}$Be which emits neutrons up to 14 MeV). This is a consequence of having achieved a fairly high flat efficiency.

\begin{figure}[htb]
\begin{center}
\includegraphics[width=0.99\linewidth]{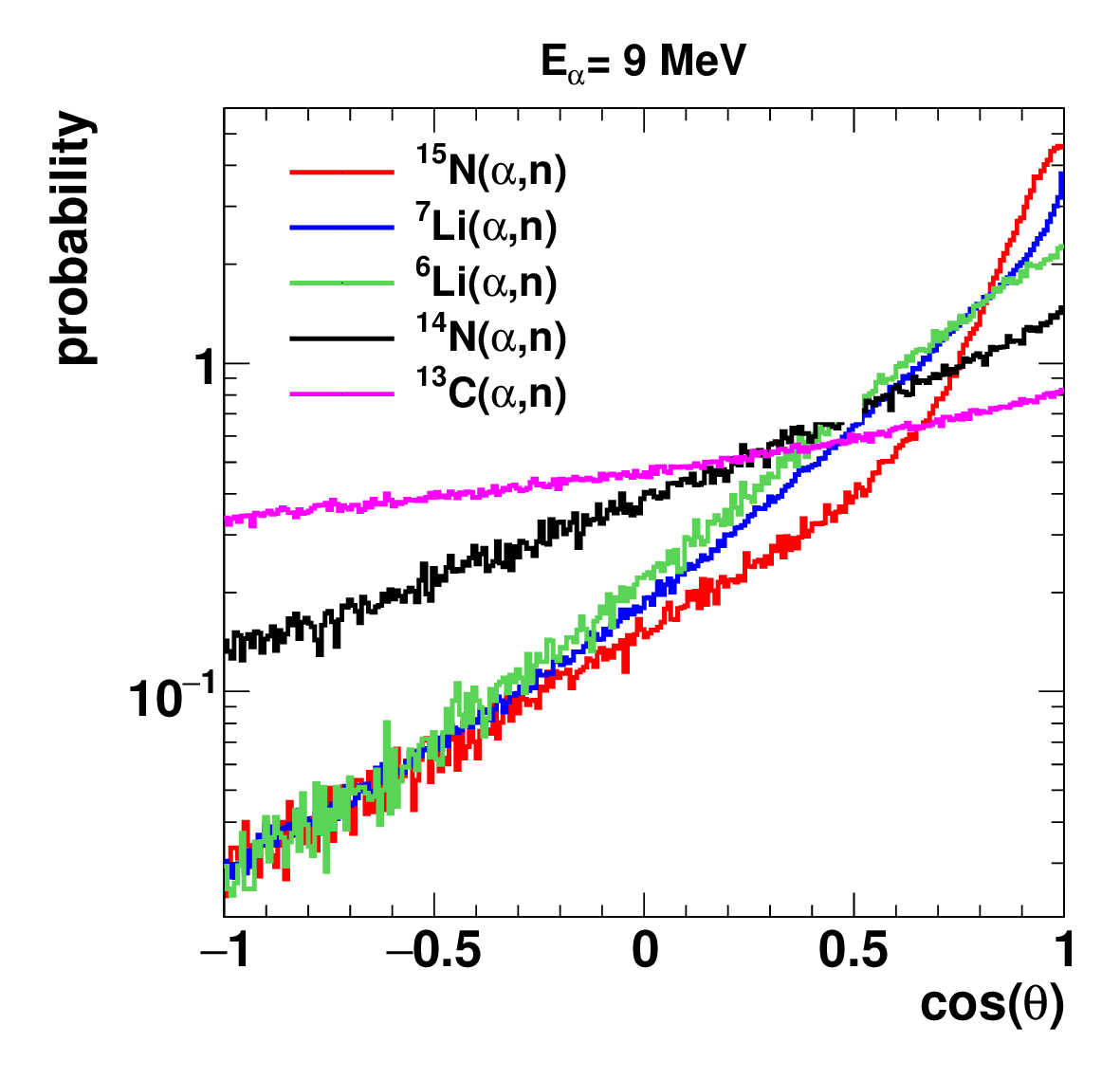} \\
\end{center}
\caption{Examples of angular distributions of the neutrons emitted in ($\alpha$,n) reactions for thick monoisotopic samples and 9 MeV incident $\alpha$-particles obtained with SaG4n using JENDL/AN-2005.}
\label{fig:AngDist01}
\end{figure}

A similar set of simulations, but with neutrons emitted according to the energy-angular distributions from SaG4n rather than isotropically, produces the red circles labeled as \textit{Nominal}. The deviations from C/R=1 are therefore due the angular distributions of the source, which differ from the isotropic distribution used to compute the response matrix $\varepsilon_{i,j}$. In this case deviations are larger than before, but remain below 2\% with three exceptions: $^{6}$Li (5\%), $^{7}$Li (7\%) and $^{15}$N (12\%). To give an idea of how anisotropic these sources are, we show some of them in Fig.~\ref{fig:AngDist01}. The anisotropy is, in fact, very high for these three cases: in an isotropic distribution, 10\% of the neutrons are emitted with $\cos(\theta)>0.8$; in the cases shown in the figure, the fraction of neutrons emitted with $\cos(\theta)>0.8$ are 60\% ($^{15}$N), 44\% ($^{7}$Li), 37\% ($^{6}$Li), 24\% ($^{14}$N), and 15\% ($^{13}$C).

Finally, we present simulations in which neutrons are emitted isotropically, with the source position shifted by 1 cm along each of the Cartesian axes: horizontal, vertical, and the beam direction (as shown in the left panel of Fig.~\ref{fig:Conf-nSPIRAL}). This serves to demonstrate nSPIRAL’s sensitivity to the source position. As can be seen in Fig.~\ref{fig:AngPosDepend01}, the detection device exhibits very low sensitivity to shifts in the source position along the beam line or in the horizontal direction (represented by the magenta crosses and green triangles, respectively). Conversely, sensitivity in the vertical direction is much higher. In the black triangles shown in the figure, the source has been simulated 1 cm below the center; that is, closer to the detectors. This increases detection efficiency and leads to an overestimation of the number of neutrons from the source by 4–5\%.

\section{Summary and conclusions} \label{sec:Conclusions}

In this work, we have demonstrated how to use the weighting technique for measurements with neutron long counters. This technique offers two fundamental advantages: (i) it enables detection efficiencies that are significantly flatter than those obtained using more conventional data analysis techniques; and (ii) it allows for greater flexibility in the geometry of the long counter. Taking advantage of this second benefit, we are proposing a long-counter design, which we have named nSPIRAL (for neutron Spectrometric Position-weIghted Radial Array Long-counter), in which almost each $^{3}$He tube is positioned at a different distance from the neutron source, with the aim of using it as a spectrometer in addition to a counter.

Besides nSPIRAL, we have presented two other designs for neutron long counters: one consisting of three rings of $^{3}$He tubes (Conf-3R), with the radii of the rings optimized; and another in which the $^{3}$He tubes are arranged in a spiral around the source (Conf-Spiral). The design of nSPIRAL is similar to the one of Conf-Spiral but with all the $^{3}$He tubes located to one side of the neutron source, rather than surrounding it. This offers three advantages: (i) the size of the long counter is reduced; (ii) access to the neutron source is significantly easier, as a cover can be constructed to allow such access; and (iii) it is straightforward to include a Pb converter for measurements in which neutrons have energies exceeding 10–15 MeV. The fundamental disadvantage of nSPIRAL compared to Conf-3R is that its detection efficiency $\varepsilon^{*}_{S}$ is $\sim$30\% smaller; however, it offers the capability for neutron spectroscopy, in addition to the advantages we have just listed.

We have also verified that when nSPIRAL operates as a spectrometer, the estimates obtained for the number of emitted neutrons are very similar to those derived using the weighting technique. Furthermore, it provides quite decent estimates of the neutron energy distribution, although, naturally, the resolution is not as fine as that achieved with time-of-flight spectrometers.

We have studied the sensitivity of nSPIRAL to the angular distributions of the source, finding that biases in the number of emitted neutrons exceeding 2\% only appear when these distributions are extremely anisotropic. It is worth noting here that the length of the $^{3}$He tubes is 60~cm, which makes nSPIRAL less dependent on neutron angular distributions than $^3$He-based neutron long counters with shorter tubes. We have also studied the sensitivity to the source position, finding that nSPIRAL is very insensitive to source displacements in the horizontal plane. In contrast, in the vertical direction, a 1~cm displacement results in a 4–5\% bias in the calculation of the number of emitted neutrons. Therefore, for measurements conducted with nSPIRAL, it is important to keep the uncertainty in the source position along the vertical axis within a few millimeters.

Finally, it should be noted that accurately knowing the response matrix of the detection system is essential for properly performing unfolding and applying the weighting technique. Since determining this matrix experimentally is practically impossible, it must be obtained through Monte Carlo simulations. Therefore, once the detection system is constructed, the Monte Carlo model will need to be validated through measurements with known neutron sources.

\section*{Acknowledgments} 

This work has been supported in part by the ENRESA-CIEMAT agreement “Tecnolog\'{i}as Disponibles para la Transmutaci\'{o}n de Radionucleidos de Vida Larga”; by projects PID2024-158174OB-I00 and PID2022-142589OB-I00, both funded by MCIN/AEI/10.13039/501100011033/FEDER, UE; and by the European Union’s HORIZON-EUROPE Research and Innovation Programme under grant agreement No. 101164596 (APRENDE, addressing priorities of evaluated nuclear data in Europe).

\FloatBarrier

\bibliographystyle{unsrt}    
\bibliography{bibliography}

\end{document}